\documentclass[12pt]{article}
\usepackage{amsfonts}
\usepackage{latexsym}
\usepackage{amsmath}
\usepackage{amssymb}

\usepackage{slashed}
\usepackage{upgreek}
\usepackage{mathrsfs}
\usepackage{bbm}

\newcommand{\newsection}{    
\setcounter{equation}{0}\section}
\def\appendix#1{\addtocounter{section}{1}\setcounter{equation}{0}
\renewcommand{\thesection}{\Alph{section}}
\section*{Appendix \thesection\protect\indent \parbox[t]{11.15cm}{#1}}
\addcontentsline{toc}{section}{Appendix \thesection\ \ \ #1}}

\newcommand{\be}{\begin{eqnarray}}
\newcommand{\ee}{\end{eqnarray}}
\newcommand{\bea}{\begin{eqnarray}}
\newcommand{\eea}{\end{eqnarray}}
\newcommand{\ba}{\begin{array}}
\newcommand{\ea}{\end{array}}
\newcommand{\nn}{\nonumber \\}

\newcommand{\la}{\label}

\newcommand{\tM}{\text{\tiny $M$}}

\newcommand{\tN}{\text{\tiny $N$}}

\newcommand{\tP}{\text{\tiny $P$}}

\def\a{\alpha}
\def\b{\beta}

\def\e{\epsilon}

\def\bbe{{\bf{e}}}
\font\mybb=msbm10 at 11pt

\def\bb#1{\hbox{\mybb#1}}

\def\bR {\bb{R}}

\def\bH {\bb{H}}
\def\bC {\bb{C}}

\def\lc{\lrcorner}
\def\ra{\rangle}

\def\ui {{\tM}}
\def\uj {{\tN}}

\def\ua{{\mathfrak {a}}}
\def\ub{{\mathfrak {b}}}

\def\uM{I}
\def\uA{{\mathfrak {i}}}
\def\uB{{\mathfrak {j}}}

\def\pr{{{r'}}}
\def\ps{{{s'}}}
\def\pt{{ {t'}}}

\def\ur{{\underline {r}}}
\def\us{{\underline {s}}}

\def\uo{{\underline {1}}}
\def\u2{{\underline {2}}}

\def\pr{{{r'}}}
\def\ps{{{s'}}}
\def\pt{{ {t'}}}

\def\ur{{\underline {r}}}
\def\us{{\underline {s}}}

\def\ka{\mathfrak{a}}
\def\kb{\mathfrak{b}}

\def\ki{\mathfrak{i}}
\def\kj{\mathfrak{j}}

\def\km{\mathfrak{m}}
\def\kn{\mathfrak{n}}
\def\kp{\mathfrak{p}}

\def\scrF{\mathscr {F}}

\begin{document}
\begin{titlepage}
\begin{center}
\vspace{5.0cm}

\vspace{3.0cm} {\Large \bf Spinorial geometry, horizons and superconformal symmetry in six dimensions}
\\
[.2cm]

{}\vspace{2.0cm}
 {\large
M.~ Akyol ~and ~G.~Papadopoulos
 }\\

{}

\vspace{1.0cm}
Department of Mathematics\\
King's College London\\
Strand\\
London WC2R 2LS, UK\\
\medskip
{\small ma483@cam.ac.uk;~~george.papadopoulos@kcl.ac.uk}
\\

\end{center}
{}
\vskip 3.0 cm
\begin{abstract}
The spinorial geometry method of solving Killing spinor equations is reviewed as it applies to 6-dimensional (1,0) supergravity. In particular, it is explained how the method is used
to identify both the fractions of supersymmetry preserved by and the geometry of all supersymmetric backgrounds. Then two applications
 are described to systems that exhibit superconformal symmetry. The first is the proof that some
  6-dimensional black hole horizons are locally isometric to $AdS_3\times \Sigma^3$, where
 $\Sigma^3$ is diffeomeorphic to $S^3$. The second one is a description of all supersymmetric solutions
of 6-dimensional (1,0)  superconformal theories and in particular of their brane solitons.
\end{abstract}

\vfill


\end{titlepage}

\setcounter{section}{0}
\setcounter{subsection}{0}


\newsection{Introduction}

The main purpose of this review article is to describe the spinorial geometry method \cite{spingeom} as it applies into the classification of
supersymmetric backgrounds of 6-dimensional (1,0) supergravity theories and then present two
applications. One application is an investigation into the geometry of black hole horizons and the other  the construction
of brane solitons in (1,0)-superconformal theories.

From the very beginning of supersymmetry theories, solutions that preserve some of the supersymmetry of an underlying theory have had
a central role in the description of their classical and quantum properties. In   supersymmetric gauge  theories, such solutions include the solitons and instantons, see eg \cite{manton} for a review,  which have applications in the understanding of these theories
at strong coupling \cite{olive, witten}. In the context of supergravity,   supersymmetric solutions either serve as backgrounds for compactifications
or describe certain classes of black hole solutions, see \cite{duff} and \cite{maeda} for reviews. These  results from supergravity theories  were later adapted to string theory and M-theory. In addition string
theory and M-theory open the arena for new classes of supersymmetric solutions, like those of branes and their intersections, see eg \cite{stelle, smith} and references within. Such new solutions have been instrumental
in the foundation of string and M-theory dualities and as well as in AdS/CFT, see eg \cite{obers, maldacena} for reviews.

Initially, the construction of supersymmetric solutions either in supergravity or in string/M-theory has been centred
around an ansatz on the fields motivated by symmetries of the physical object or process under investigation. Such an approach has been very successful and
has produced a vast number of solutions many of which have some key applications. However, such an approach is rather limiting
as it focuses on particular points of what may be a large set, or moduli space, of similar solutions and it lacks an overview of the possibilities
that may be available. Therefore to gain an insight into the structure of string theory and M-theory as well as for many applications
in AdS/CFT and black holes a more systematic approach to the construction of supersymmetric solutions is needed.

The problem of classifying the supersymmetric solutions of supergravity theories has been known for sometime. Using twistor methods, P Tod classified
the supersymmetric solutions of simple 4-dimensional supergravities \cite{tod}. Gauntlett et al in \cite{hull} solved  KSEs of minimal 5-dimensional supergravity using  a technique based
 on  spinor bi-linears, and later this was applied in \cite{pakis} to solve the Killing spinor equations (KSEs) of D=11 supergravity for one spinor.   J Figueroa-O'Farrill and one of the authors classified the maximal supersymmetric solutions
of 10- and 11-dimensional supergravities   using the integrability conditions of the KSEs  \cite{gpjose}.

The spinorial geometry method proposed Gillard, Gran and one of the authors in \cite{spingeom}  utilizes spinorial techniques and it was originally applied
 to solve the KSEs of 11-dimensional supergravity. One of its characteristics is that it provides a systematic
  way to solve the KSEs of supersymmetric theories \cite{systway}. The main results of this method have been the solution of the
KSEs of IIB and IIA supergravities for one Killing spinor \cite{iibgrangp, iiagrangp}, and the solution of KSEs of heterotic \cite{het1, het2} and 6-dimensional (1,0) supergravities \cite{ap1} in all cases, as well as many other applications in other lower dimensional supergravity theories, see eg \cite{grover1, d4grangp, klemm1, grover2, klemm2}.
In addition, spinorial geometry has been used to classify all near maximal supersymmetric solutions of IIB \cite{iibnearmax} and 11-dimensional supergravities \cite{d11nearmax}.

The spinorial geometry method to solving KSEs  is based on three ingredients. To describe these ingredients, first observe that all supergravity theories
have a local gauge group which includes  $Spin(D)$ as  a subgroup, where $D$ is the dimension of spacetime. The first ingredient of spinorial geometry is
to use the gauge group of a supergravity theory to locally choose representatives of the Killing spinors.  These are labeled by the orbits of the
gauge group on the space of spinors. The second ingredient is  a realization of spinors in
terms of forms which simplifies the way that the KSEs act on the spinor representatives,
 and the third the use of an oscillator basis in the space of spinors
which allows the rewriting of the KSEs in terms of a linear system. This linear system has as
unknowns components of the fluxes as well as components of the spin connection
of the supergravity theory. The linear system is then solved to express some of the fluxes in
terms of the geometry and also find the restrictions on the
geometry required for the existence of Killing spinors. The latter restrictions are typically
expressed as a linear relation between the components
of the spin connection. The expressions of the fluxes in terms of the geometry and the conditions on the geometry can be organized in
irreducible representations of the isotropy group of the Killing spinors in the gauge group of the supergravity theory.

For the application of spinorial geometry at hand, we shall describe how the spinorial geometry
 has been used  in \cite{ap1} to solve the KSEs of 6-dimensional
(1,0) supergravity coupled to any number of vector, tensor and  scalar multiplets \cite{sezgin, ferrara, riccioni} in all cases.
In particular, we shall describe how all the fractions of supersymmetry
preserved by the backgrounds have been identified as well as what is the geometry of the underlying spacetime in all cases.

Furthermore, we shall present two applications of the above results in the context of
superconformal systems. One application is the classification of all near horizon geometries of 6-dimensional
(1,0) supergravity coupled to tensor and scalar multiplets described in \cite{ap2}. In particular, we shall show that a class of  horizons is isometric to
$AdS_3\times \Sigma^3$, where the universal cover of $\Sigma^3$ is diffeomorphic to $S^3$,
and depending on the geometry of $\Sigma^3$ can preserve 2, 4 or 8 supersymmetries.

Another application that we shall demonstrate is the solution of the KSEs of (1,0)-superconformal theories  in 6 dimensions \cite{ap3, ap4}. Such theories have been proposed, \cite{ssw, ssw3},  in the context
of finding a Lagrangian description for a multiple M5-brane theory  which is conjectured to be the field theory dual of M-theory on $AdS_7\times S^4$.
We shall demonstrate that large classes of such (1,0)-superconformal symmetries have soliton solutions which are expected from the M-brane intersection rules \cite{strominger, pktgp}.

\newsection{Spinorial geometry}
\label{paradigm}
\subsection{A paradigm}

Before we proceed to apply the spinorial geometry method to solve the KSEs of 6-dimensional supergravity, we shall illustrate
how this works in an example.  For this consider the KSE
\bea
F_{\mu\nu} \Gamma^{\mu\nu} \epsilon=0~,
\la{kse6d}
\eea
which arises in 6-dimensional Euclidean gauge theory, where $\epsilon$ is a spinor, $F$ is a gauge field strength on $\bR^6$ and the gauge indices are suppressed.
Solution of this equation means to find the geometric conditions on $F$ such that there is an $\epsilon\not=0$, called Killing spinor, which
solves the above equation.

As we have mentioned the spinorial geometry method proceeds in  three steps. First is to identified the orbits of the gauge group
of the system on the space of spinors, second is to realize the spinors in terms of forms, and third is to use a basis in the
space of spinors to turn the KSEs into a linear system. This system then   can be solved find the conditions of $F$ such that
(\ref{kse6d}) has a solution. In practise all steps are related as if one has a convenient realization of spinors as in step 2, then
it is more convenient  to find the orbits of the gauge group on the space of spinors required in step 1, and to introduce a basis so that step 3 can be carried out.
So let us begin with step 2.

\subsection{Spinors in terms of forms}

Let us consider the spinor representations of $Spin(6)$. These can be constructed by identifying the Dirac representation with the space of forms on $\bC^3$, $\Lambda^*(\bC^3)$.
Then a realization of Dirac gamma matrices is
\bea
\Gamma_i= e_i\wedge + e_i\lc~,~~~\Gamma_{3+i}=i \big(e_i\wedge - e_i\lc\big)~,~~~i=1,2,3~,
\eea
where $(e_i)$ is a Hermitian basis in $\bC^3$ and $\lc$ is the inner derivation operation on $\Lambda^*(\bC^3)$ which is adjoint to the wedging.
One can verify the that above gamma matrices $(\Gamma_a)=(\Gamma_i, \Gamma_{3+i})$ satisfy the Clifford algebra relations $\Gamma_a \Gamma_b+\Gamma_b \Gamma_a=2 \delta_{ab}$.

The decomposition of forms in even and odd according to their degree, $\Lambda^*(\bC^3)= \Lambda^{\rm ev}(\bC^3)\oplus \Lambda^{\rm od}(\bC^3)$,
corresponds to the decomposition of the Dirac representation into chiral (Weyl) and anti-chiral (anti-Weyl) representations.  The Dirac and chiral representations are complex.
There is a real (Majorana) representation of $Spin(6)$ as well identified as the  eigenspace of the  operator, $R=\Gamma_{456} *$,  in $\Lambda^*(\bC^3)$
with eigenvalue 1. Observe that $R^2=1$ and that $R$ is anti-linear. Real (Majorana) spinors have both chiral and anti-chiral components.

\subsection{Orbits of the gauge group and linear system}

To identify the gauge group of KSE (\ref{kse6d}), observe that under a $Spin(6)$ transformation of $\epsilon$, the  KSE  transforms covariantly provided that there is a compensating $SO(6)$ rotation of $F$. Therefore
 the gauge group of the KSE (\ref{kse6d}) is $Spin(6)$. Since the solutions $\epsilon$ of the KSE are identified
 up to a gauge transformation, the independent solutions are labeled by the orbits of the gauge group in the
 space of spinors or the orbits of the gauge group in appropriate number of tensor copies for more than one Killing spinor. For the solution of the KSE, any representative of $\epsilon$ in an orbit can be chosen.

 To find the orbits of $Spin(6)$ in the space of spinors, it is convenient to use the isomorphism $Spin(6)=SU(4)$. Under this isomorphism,  the chiral and anti-chiral representations of $Spin(6)$
are identified with the fundamental and anti-fundamental representations ${\bf 4}$ and $\bar{ \bf 4}$ of $SU(4)$, respectively. As a result
$Spin(6)=SU(4)$ has one type of a non-trivial orbit in each of these two representations  which is a 7-sphere and has isotropy group $SU(3)$. Therefore assuming that $\epsilon$ is chiral or anti-chiral,  it can be put in any direction in the  ${\bf 4}$ or $\bar{ \bf 4}$ representation, respectively.

To solve the KSE, it is convenient to choose a ``simple'' representative for the Killing spinor. To do this assume that $\epsilon$ is chiral and observe that
\bea
\Lambda^{\rm ev}(\bC^3)= \bC\langle 1, e_{ij}\rangle~,
\eea
where $e_{ij}=e_i\wedge e_j$.
As $\epsilon$ can be put in any direction, one can choose without loss of generality that $\epsilon=1$. Then the KSE (\ref{kse6d}) can be rewritten as
\bea
F_{\mu\nu} \Gamma^{\mu\nu} 1=0~.
\la{kse6d1}
\eea

To find the linear system associated to the above equation, introduce a Hermitian basis in the space of gamma matrices as
\bea
\Gamma_\alpha={1\over \sqrt 2} (\Gamma_\alpha-i \Gamma_{\alpha+3})~,~~~\Gamma_{\bar \alpha}={1\over \sqrt 2} (\Gamma_\alpha+i \Gamma_{\alpha+3})~,
\eea
where now
\bea
\Gamma_\alpha\Gamma_\beta+ \Gamma_\beta \Gamma_\alpha=0~,~~~\Gamma_{\bar\alpha}\Gamma_{\bar\beta}+ \Gamma_{\bar\beta} \Gamma_{\bar\alpha}=0~,~~~\Gamma_\alpha\Gamma_{\bar\beta}+ \Gamma_{\bar\beta} \Gamma_\alpha=2 \delta_{\alpha\bar\beta}~.
\eea
Expanding (\ref{kse6d1}) in this Hermitian basis, one finds the linear system
\bea
F_{\bar\alpha\bar\beta} \Gamma^{\bar\alpha\bar\beta} 1+ 2 \delta^{\alpha\bar\beta} F_{\alpha\bar\beta} 1=0~.
\eea
Since $(1, \Gamma^{\bar\alpha\bar\beta} 1)$ for $\bar\alpha<\bar\beta$ is a basis in $\Lambda^{\rm ev}(\bC^3)$, we conclude that the solution to the linear system is
\bea
F_{\bar\alpha\bar\beta}=0~,~~~\delta^{\alpha\bar\beta} F_{\alpha\bar\beta}=0~.
\label{fcon}
\eea
To interpret  these conditions,  one can define a 2-form spinor bilinear as
\bea
\omega={i\over2} \langle 1, \Gamma_{ij} 1\rangle \,dx^i\wedge dx^j~,
\eea
where $\langle \cdot, \cdot \rangle$ is the Dirac spinor inner product which is  the same as the Hermitian inner
product on the space of spinors $\Lambda^*(\bC^3)$ induced from that on $\bC^3$. This is a Hermitian form and together with the metric on $\bR^6$  give rise
to a complex structure on $\bR^6$. This is the complex structure $I$ which is invariant under the isotropy group $SU(3)$ of the Killing spinor. Therefore in complex coordinates on $\bR^6$ with respect to $I$
\bea
ds^2=\delta_{ij} dx^i dx^j=2 \delta_{\alpha\bar\beta}\,dz^\alpha dz^{\bar \beta}~,~~~\omega=-i \delta_{\alpha\bar\beta} dz^\alpha\wedge dz^{\bar\beta}~.
\eea
Then the above conditions (\ref{fcon})
 imply that $F$ is a (2,0) and (1,1) form with respect to $I$, and the trace of the (1,1) component vanishes. Of course if $F$ is real,
 then the (2,0) component vanishes as well as it is the complex conjugate of (0,2) component. These conditions can immediately be recognized as  instanton equations  on 6-dimensions refereed to as the Hermitian-Einstein conditions on the gauge fields.

 \subsection{Spinorial geometry and supergravity}

 The KSEs of supergravity theories are the vanishing conditions of the supersymmetry variations of the fermions
 of the theory evaluated at the locus where all fermions vanish. The unknowns are the supersymmetry parameters which are taken
 to be commuting spinors. The KSEs of supergravity theories are separated into parallel transport equations associated with the supersymmetry variations
 of the gravitini, and algebraic equations associated with the supersymmetry variations of the remaining
 fermions of the theory.  Schematically, they are written as
 \bea
 {\cal D}_\mu \epsilon &\equiv& \nabla_\mu \epsilon+ \Sigma_\mu(g, F)\, \epsilon =0~,
 \cr
 {\cal A}\,\epsilon&\equiv&{\cal A}(g, F)\, \epsilon=0~,
 \eea
 where $\epsilon$ is the supersymmetry parameter.
 The covariant derivative, ${\cal D}$, often called the supercovariant derivative of the supergravity theory,
 begins with the spin connection of the spacetime metric, $\nabla$, and receives a correction $\Sigma$ which depends on the metric and the remaining
 bosonic fields of the theory. $\Sigma$ typically contains terms of
 higher order than quadratic in a shew-symmetric product expansion of gamma matrices.  On the other hand ${\cal A}$ is an algebraic equation on $\epsilon$ which depends on the bosonic
 fields of the theory.

 The gauge transformations of  KSEs are those transformations which leave the form of the KSEs covariant. The gauge group of the KSEs of a supergravity theory  includes the spin group of the spacetime, and in gauged supergravity
 also includes the gauge group of the theory. The holonomy group of the supercovariant
 connection ${\cal D}$ of a generic background includes the gauge group of the KSEs and  but in many supergravity theories
 is a much larger  group.

 In the current context, a solution of the KSEs  means to specify the differential geometric conditions on the bosonic fields of the supergravity
 theory such that the KSEs admit an $\epsilon\not=0$ as a solution. The number $N$ of linearly independent solutions  $\epsilon$, called Killing spinors, of
 the KSEs is the number of supersymmetries preserved by the background. To find a supersymmetric solution in addition to solving the
 KSEs, one also has  to solve the field equations of the theory. Typically the KSEs imply some of the field equations but not necessarily all.

 Spinorial geometry utilizes the gauge group of the KSEs of a supergravity theory to choose the Killing spinors. Then
 as in  the gauge theory paradigm, the KSEs turn into a linear system which is solved to express some of the
 fields in terms of the geometry and also identify the conditions on the geometry required for the KSEs to admit a solution.
 The geometric conditions are typically expressed as a linear relation between the components of the
 spin connection $\nabla$.

\newsection{$(1,0)$ supergravity in six dimensions}

The main task is to describe the solution of the KSEs of 6-dimensional (1,0) supergravity  coupled to any number
of tensor, vector and scalar multiplets  presented in \cite{ap1}.
Solutions of the KSEs of 6-dimensional supergravities in special cases have been investigated before in \cite{dario, jose,
han, jong, gueven}.

\subsection{Fields and KSEs}

Supergravity in six dimensions \cite{sezgin, ferrara, riccioni} with (1,0) supersymmetry, 8 real supercharges, is constructed from  four different supersymmetry multiplets  the following. The gravitational multiplet which has field content
 a graviton $g$, an anti-self-dual 2-form gauge potential $B$ and a gravitino $\Psi$.  The tensor multiple which consists of a self-dual 2-form gauge potential $b$, a scalar $\phi$ and fermion $\chi$
 which has chirality opposite to that of the gravitino. The vector or gauge multiplet which has a vector gauge potential $A$ and a fermion $\lambda$
 with the same chirality as that of  gravitino, and a scalar or hyper-multiplet which consists of four real scalars $q$ and
 a fermion $\psi$ which has opposite chirality to that of  gravitino. A mnemonic of the field content of the
 multiplets is

\bea
\mathrm {gravity~~ multiplet}&:&~~g_{\mu\nu}~,~~~B_{\mu\nu}~;~~~\Psi_\mu
\cr
\mathrm {tensor~~ multiplet}&:&~~b_{\mu\nu}~,~~~\phi~;~~~\chi
\cr
\mathrm {vector~~ multiplet}&:&~~A_\mu~;~~~\lambda
\cr
\mathrm {scalar~~ multiplet}&:&~~q~;~~~\psi
\label{mne}
\eea

 The system that we consider is (1,0) supergravity coupled to $n_T$
 tensor, $n_V$ vector and $n_H$ scalar multiplets. All the fermions of the four multiplets are chiral  and satisfy the symplectic-Majorana spinor condition. The symplectic-Majorana condition is a reality condition which is imposed on the complex chiral spinors of $Spin(5,1)$.  This condition utilizes the invariant $Sp(1)$ and $Sp(n_H)$ forms
to impose a reality condition on the complex spinors preserving chirality. Suppose that the Dirac or Weyl spinors $\lambda$ and $\chi$ transform under the
fundamental representations of $Sp(1)$ and $Sp(n_H)$, respectively. The symplectic Majorana condition is given by
\bea
\lambda^\uA= \epsilon^{\uA\uB} C \bar\lambda^T_\uB~,~~~\chi^\ua= \epsilon^{\ua\ub} C \bar \chi^T_\ub~,
\eea
where $C$ is the charge conjugation matrix and $\epsilon^{\uA\uB}$ and $\epsilon^{\ua\ub}$ are the symplectic invariant
forms of $Sp(1)$ and $Sp(n_H)$, respectively, and $\uA, \uB=1,2$ and $\ua,\ub=1,\dots, 2n_H$.

To describe the KSEs of (1,0) supergravity coupled to tensor, vector and scalar multiplets,
we use a formulation\footnote{We use a different normalization for some of the fields from that in \cite{riccioni}. Our normalization
is similar to that of heterotic supergravity.}  proposed by \cite{riccioni}.
The theory has $n_T+1$ 2-form gauge potentials
$B^\ur$, $\ur=0,1,\dots, n_T$.
One of the 2-form potentials is associated  with the gravitational multiplet and the remaining  $n_T$ with the tensor multiplets.
Let us denote the corresponding 3-form field strengths with $G^\ur$.
To continue, the scalar fields of the tensor multiplets parameterize the coset space $SO(1,n_T)/SO(n_T)$.
 A convenient way to describe this coset space is to choose a local section  $S$  as
\begin{equation}
S= \begin{pmatrix}
v_\ur\\ x_\ur^\uM
 \end{pmatrix}~,~~~\uM=1,\dots n_T
\end{equation}
Since $S\in SO(1, n_T)$, one has $\tilde S\eta S= \eta$ where $\eta$ is the Lorentz metric in $n_T+1$-dimensions. In
particular
\bea
v_\ur v^\ur=1~,~~~v_\ur v_\us- \sum_\uM x_\ur^\uM x_\us^\uM=\eta_{\ur\us}~,~~~v^\ur x_\ur^\uM=0~.
\eea

The scalars of the hypermultiplet parameterize a Quaternionic K\"ahler manifold ${\cal Q}$. This is a Riemannian manifold
 equipped with a quaternionic structure, ie endomorphisms $I_\tau$, $\tau=,1,2,3$, of the tangent bundle such that
 $I_{\tau_1} I_{\tau_2}=- \delta_{\tau_1\tau_2} {\bf 1}+ \epsilon_{\tau_1\tau_2\tau_3} I_{\tau_3}$, and whose Levi-Civita connection  has holonomy
$Sp(n_H)\cdot Sp(1)$, see \cite{salamon} for a mathematical description. Such a manifold admits a frame $E$ such that the metric and the endomorphisms can be written as
\bea
g_{\ui\uj}= E_\ui^{\ua \uA} E_\uj^{\ub \uB} \epsilon_{\ua\ub} \epsilon_{\uA\uB}~,~~~(I_\tau)^\tM{}_\tN=-i (\sigma_\tau)^\ki{}_\kj
\delta^\ka{}_\kb E_{\ua \ki}^\tM E^{\ub \kj}_\tN~,
\eea
where $\epsilon_{\ua\ub}$ and $\epsilon_{\uA\uB}$ are the invariant $Sp(n_H)$ and $Sp(1)$ 2-forms, respectively, and $\sigma_\tau$ are the Pauli matrices. The
spin connection, which has holonomy $Sp(n_H)\cdot Sp(1)$,  decomposes as ${\cal A}_\tM=({\cal A}_\ui^\ua{}_\ub, {\cal A}_\ui^\uA{}_\uB)$.

In \cite{riccioni} to include vector multiplets with (non-abelian) gauge potential $A_\mu^\km$, one assumes that the
Quaternionic K\"ahler manifold\footnote{It is likely that this assumption is not necessary and a more general class
of models can exist. Moreover $\mu$ may be related to moment maps \cite{gal} of Quaternionic K\"ahler geometry.} ${\cal Q}$ of the hypermultiplet
is $Sp(n_H,1)/Sp(1)\times Sp(n_H)$ and gauges the maximal
compact isometry subgroup $Sp(1)\times Sp(n_H)$. So the gauge group of the theory is $H=Sp(1)\times Sp(n_H)\times K$, where $K$
is a product of semi-simple groups which does not act on the scalars. Let $\xi_{\km_1}$ and $\xi_{\km_2}$ be the vector fields
generated on $Sp(n_H,1)/Sp(1)\times Sp(n_H)$ by the action of $Sp(1)$ and $Sp(n_H)$, respectively.
 Under these assumptions, one defines
\bea
&&H_{\mu\nu\rho}=v_\ur G^\ur_{\mu\nu\rho}~,~~~H^\uM_{\mu\nu\rho}=x^\uM_\ur G^\ur_{\mu\nu\rho}~,~~~
{\cal C_\mu}{}^\uA{}_\uB=D_\mu q^{\ui} {\cal A}_\ui{}^\uA{}_\uB~,
\cr
&&T^\uM_\mu=x_\ur^\uM\partial_\mu v^\ur~,~~~ V^{\ua\uA}_\mu=E^{\ua\uA}_\ui D_\mu q^\ui~,~~~
F^{\km}_{\mu\nu}= \partial_\mu A_\nu^{\km}-\partial_\nu A_\mu^{\km}+ f^{\km}{}_{\kn\kp} A_\mu^{\kn} A_\nu^{\kp}~,
\cr
&&(\mu^{\km_1})^\uA{}_\uB  = -{2\over v_\ur c^{\ur 1}}
{\cal A}_\ui{}^\uA{}_\uB \xi^{\ui \km_1}~,~~~(\mu^{\km_2})^\uA{}_\uB  = -{2\over v_\ur c^{\ur 2}}
{\cal A}_\ui{}^\uA{}_\uB \xi^{\ui \km_2}~,~~
(\mu^{\km_3})^\uA{}_\uB  =0~,
\la{ric}
\eea
where the gauge index $\km_3$ ranges over the gauge subgroup $K$,  $q^\ui$ are the scalars of the hypermultiplet,
\bea
\nabla_\mu\epsilon^\uA=\partial_\mu\epsilon^\uA+{1\over4} \Omega_{\mu, mn} \gamma^{mn} \epsilon^\uA~,~~
D_\mu q^\ui=\partial_\mu q^\ui- A_\mu^{\km} \xi_{\km}^\ui~,
\eea
 and $\Omega$ is the frame connection of spacetime. It is understood that $\xi_{\km_3}=0$
as $K$ does not act on the scalars of the hypermultiplet.
Clearly $F^{\km}$ are the field strengths of the gauge potentials $A^\km$
and $f$ are the structure constants of the gauge group $H$. We refer to  $\mu$'s as the moment maps, see  \cite{gal}.

It remains to define the field strengths $G^\ur$. These are given by
\bea
G^\ur_{\mu\nu\rho} =3 \partial_{[\mu} B^\ur_{\nu\rho]}+ c^{\ur 1} CS(A^{Sp(1)})_{\mu\nu\rho}+ c^{\ur 2}
CS(A^{Sp(n_H)})_{\mu\nu\rho}+ c^{\ur K} CS(A^{K})_{\mu\nu\rho}~,
\eea
where $c^\ur$\,'s are constants, one for each copy of the gauge group, and $CS(A)$'s are the Chern-Simons 3-forms.
Observe that the constants $c^{\ur 1}$ and $c^{\ur 2}$
enter in the definition of $\mu$'s in (\ref{ric}).

The duality condition on $G$ is given by
\bea
\zeta_{\ur\us} G^\us_{\mu_1\mu_2\mu_3}={1\over 3!} \epsilon_{\mu_1\mu_2\mu_3}{}^{\nu_1\nu_2\nu_3} G_{\ur \nu_1\nu_2\nu_3}~,
\eea
where
\bea
\zeta_{\ur\us}= v_\ur v_\us+ \sum_\uM x^\uM_\ur x^\uM_\us~.
\eea
Note that the duality conditions for $H$ and $H^\uM$ are opposite. In our conventions, $H$ is anti-self-dual while $H^\uM$
are self-dual.

The Lagrangian of the theory is
\bea
e^{-1}\mathcal{L} & = & -\frac{1}{4}R+\frac{1}{48}\varsigma_{\ur\us}G_{\mu\nu\rho}^{\ur}G^{\us\ \mu\nu\rho}-\frac{1}{4}\partial_{\mu}v^{\ur}\partial^{\mu}v_{\ur}+\frac{1}{8}v_{\ur}c^{\ur}F_{\mu\nu}^{\km}F^{\km\mu\nu}
\cr
 &  & -\frac{1}{64e}\epsilon^{\mu\nu\rho\sigma\delta\tau}
 B_{\mu\nu}^{\ur}c_{\ur}F_{\rho\sigma}^{\km}
 F_{\delta\tau}^{\km}+\frac{1}{2}g_{\ui\uj}D_{\mu}q^{\ui}D^{\mu}q^{\uj}
\cr
&&
 -\frac{1}{2v_{\ur}c^{\ur}}\mathcal{A}_{\ui\ \ki}^{\kj}\mathcal{A}^{\ki}_{\uj\ \kj}\,\xi^{\km\ui}\xi^{\km \uj}~.
\label{lagrangian}
\eea
It is understood that to derive the field equations one first varies  3-form field strengths
and then imposes  self-duality  and anti-self-duality conditions.

The  supersymmetry transformations of (1,0) supergravity fermions coupled to $n_T$ tensor, $n_V$ vector and $n_H$ scalar multiplets   evaluated at the locus where all the fermion fields vanish are
\bea
\delta\Psi^\uA_\mu&=&\nabla_\mu\epsilon^\uA-{1\over8}  H_{\mu\nu\rho} \gamma^{\nu\rho}\epsilon^\uA
+{\cal C}_\mu{}^\uA{}_\uB\, \epsilon^\uB~,
\cr
\delta \lambda^{\km\uA}&=& -{1\over 2\sqrt 2} F^{\km}_{\mu\nu} \gamma^{\mu\nu} \epsilon^\uA-{1\over\sqrt{2}} (\mu^{\km})^\uA{}_\uB  \e^\uB~,
\cr
\delta \chi^{\uM\uA}&= &{i\over2} T^\uM_\mu \gamma^\mu \epsilon^\uA
-{i\over24} H^\uM_{\mu\nu\rho} \gamma^{\mu\nu\rho} \epsilon^\uA~,
\cr
\delta \psi^\ua&=& i\gamma^\mu \epsilon_\uA  V^{\ua\uA}_\mu ~,
\la{6kse}
\eea
where the fermions are defined as in (\ref{mne}).
The KSEs of the (1,0) supergravity are derived from setting all the above transformations to zero and they will be referred to as gravitino,
gaugini, tensorini and hyperini   KSEs, respectively. Although to write the above KSEs
we have used the particular supergravity theory described in \cite{riccioni}, the form of these transformations
is model independent. The reason is that these transformations are the most
general supersymmetry transformations that one can write. So although the expression of the field strengths in terms of
the gauge potentials
will change from model to model depending on the details of the couplings, the actual form of the transformations does not.
The application of the spinorial geometry method to solve the KSEs does not depend of the details on how the field strengths
depend on the physical fields.  As a results it applies to all (1,0) supersymmetric models and not only to the one
described in this section.

\subsection{A realization of spinors in terms of forms}
\label{laforms}

The most effective way to represent the spinors of (1,0) supergravity in terms of forms is to identify the
symplectic Majorana-Weyl spinors of $Spin(5,1)$ with the $SU(2)$ invariant  Majorana-Weyl spinors of  $Spin(9,1)$ \cite{het1, ap1}.
 To do this explicitly,   the Dirac spinors
of $Spin(9,1)$ are identified with $\Lambda^*(\bC^5)$, and the positive and negative chirality
spinors are the even and odd degree forms, respectively. A realization of the gamma matrices of ${\rm Clif}(\bR^{9,1})$ is given by
\bea
\Gamma_0&=&-e_5\wedge +e_5\lc~,~~~\Gamma_5=e_5\wedge +e_5\lc~,
\cr
\Gamma_i&=&e_i\wedge +e_i\lc~,~~~\Gamma_{i+5}=i(e_i\wedge -e_i\lc)~,~~~i=1,2,3,4~,
\eea
where $e_i$, $i=1,\dots,5$, is a Hermitian basis in $\bC^5$.
The gamma matrices of ${\rm Clif}(\bR^{5,1})$ are identified as
\bea
\gamma_\mu=\Gamma_\mu~,~~~\mu=0,1,2~;~~~~ \gamma_\mu=\Gamma_{\mu+2}~,~~~\mu=3,4,5~.
\label{gammam}
\eea
Therefore the positive chirality Weyl spinors of $Spin(5,1)=SL(2,\bH)$ are
$\Lambda^{\rm ev}(\bC\langle e_1, e_2, e_5\ra)=\bH^2$.
The symplectic Majorana-Weyl condition of $Spin(5,1)$ is  the Majorana-Weyl condition
of $Spin(9,1)$ spinors, ie
\bea
\epsilon^*=\Gamma_{67} \Gamma_{89} \epsilon~,
\eea
where $\epsilon \in \Lambda^{\rm ev} \bC\langle e_1, e_2, e_5\ra\otimes \Lambda^*\bC\langle e_{34}\ra$. In particular a basis for the
symplectic Majorana-Weyl spinors is
\bea
&&1+e_{1234}~,~~~i(1-e_{1234})~,~~~e_{12}- e_{34}~,~~~i(e_{12}+ e_{34})~,~~~
\cr
&&e_{15}+e_{2534}~,~~~i(e_{15}-e_{2534})~,~~~e_{25}-e_{1534}~,~~~i(e_{25}+e_{1534})~.~~~
\label{smw}
\eea
Observe that the above basis selects the diagonal of two copies of the Weyl representation of $Spin(5,1)$, where
the first copy is $\Lambda^{\rm ev}(\bC\langle e_1, e_2, e_5\ra)$ while the second copy is $\Lambda^{\rm ev}(\bC\langle e_1, e_2, e_5\ra) \otimes \bC\langle e_{34}\ra$. The $SU(2)$ acting on the auxiliary directions $e_3$ and $e_4$ leaves the basis invariant.

The KSEs of 6-dimensional supergravity can be rewritten in terms of the 10-dimensional
notation we have introduced above. For this, we define $\rho^{\pr}$, $\pr=1,2,3$, such that
\bea
\rho^{1} =  \frac{1}{2}(\Gamma_{38}+\Gamma_{49})~,~~~\rho^{2}= \frac{1}{2}(\Gamma_{89}-\Gamma_{34})~,~~~
\rho^{3}  =  \frac{1}{2}(\Gamma_{39}-\Gamma_{48})~.
\label{spgen}
\eea
Observe that these are the generators of the Lie algebra $Sp(1)$ as it acts on the basis (\ref{smw}). Using this
the KSEs can be rewritten as
\bea
{\cal D}_\mu\epsilon\equiv \big(\nabla_{\mu}-{1\over8} H_{\mu\nu\rho} \gamma^{\nu\rho}+ {\cal C}_\mu^{\pr} \rho_{\pr}\big)\epsilon & = & 0,
\cr
\left({1\over4}F_{\mu\nu}^{\km}\gamma^{\mu\nu}+{1\over2}\mu_{\pr}^{\km}\rho^{\pr}\right)\epsilon
& = & 0~,
\cr
\left({i\over2} T^\uM_\mu \gamma^\mu-\frac{i}{24}H_{\mu\nu\rho}^{\uM}\gamma^{\mu\nu\rho}\right)\epsilon & = & 0,
\cr
i\gamma^\mu \epsilon_\uA  V^{\ua\uA}_\mu &=&0~.
\la{kkk}
\eea
In the hyperini KSE, it is understood that
\bea
\epsilon_1=-\epsilon^2~,~~~\epsilon_2=\Gamma_{34} \epsilon^1~,
\label{hypdef}
\eea
where $\epsilon^1$ and $\epsilon^2$ are the components of $\epsilon$ in the two copies of the Weyl representation used
to construct the symplectic Majorana-Weyl representation as explained below (\ref{smw}).

\newsection{Solution of KSEs}

To solve the KSEs of a supergravity theory, it is customary to begin with the gravitino KSE. This is because
 it is a parallel transport equation and so has a significant role in the description of
 the geometry of spacetime. As a result, we shall  present a  detailed
analysis of the solutions of gravitino KSE of (1,0) supergravity.   The solution of the remaining KSEs will be presented in some detail for backgrounds preserving one supersymmetry.  For the rest of the cases, only a  brief summary will be given. The omitted  details and the proof
of the statements we have used to solve all KSEs can be found in the original paper \cite{ap1}.

\subsection{Gravitino KSE}

To solve  KSEs in the context of spinorial geometry, the main task is to find the representatives of the Killing spinors
up to  gauge transformations. The gauge group of the KSEs of (1,0) supergravity is $Spin(5,1)\cdot Sp(1)$. This is the same
as the (reduced) holonomy group of the supercovariant connection ${\cal D}$ in  (\ref{kkk}) for a generic background.  To see the latter, the curvature
${\cal R}$ of the supercovariant connection is
\bea
{\cal R}_{\mu\nu}\equiv [{\cal D}_\mu, {\cal D}_\nu]={1\over4} \hat R_{\mu\nu, \rho\sigma} \gamma^{\rho\sigma} +
\scrF_{\mu\nu}^{\pr} \rho_{\pr}~,
\la{intcon}
\eea
where
\bea
\scrF_{\mu\nu}^{\pr}=\partial_\mu {\cal C}^\pr_\nu-\partial_\nu{\cal C}^\pr_\mu+2 \epsilon^{\pr}{}_{\ps\pt} {\cal C}_\mu^{\ps}
{\cal C}_\nu^{\pt}~,
\eea
and $\hat R$ is the curvature of the connection, $\hat\nabla$, with skew-symmetric torsion $H$ defined as
\bea
\hat\nabla_\mu Y^\nu=\nabla_\mu Y^\nu+{1\over2} H^\nu{}_{\mu\lambda} Y^\lambda~.
\eea
For any two vectors $X,Y$ of spacetime, ${\cal R}(X,Y)$  spans a $\mathfrak{spin}(5,1)\oplus \mathfrak{sp}(1)$ algebra
and so the holonomy of ${\cal D}$ is contained in $Spin(5,1)\cdot Sp(1)$.

Now the solutions $\epsilon\not=0$ of the gravitino KSE, ${\cal D}_\mu\epsilon=0$,  must satisfy ${\cal R}\epsilon=0$.  Thus either
the Killing spinors $\epsilon$ have a trivial isotropy group in the generic holonomy group $Spin(5)\cdot Sp(1)$ in which case
\bea
\hat R=0~,~~~~\scrF=0~.
\eea
and so the spacetime is parallelizable with
respect to a connection with skew-symmetric torsion, or they have a non-trivial  isotropy group in the generic holonomy group $Spin(5)\cdot Sp(1)$.
In the former case, all such spacetimes are locally isometric to group manifolds with anti-self-dual structure constants.
In the latter case, the holonomy of the supercovariant ${\cal D}$ connection reduces to that of the isotropy group of the Killing spinors.
 So to complete the solution of the  gravitino KSE, the subgroups of $Spin(5,1)\cdot Sp(1)$  which leave spinors invariant must be identified.

\subsubsection{Non-trivial isotropy groups}

To find the isotropy groups of spinors, it is known that the action of $Spin(5,1)\cdot Sp(1)$ on the space of symplectic Majorana-Weyl spinors can be described in terms of quaternions. In particular, the chiral symplectic Majorana spinors are identified with $\bH^2$ and  $Spin(5,1)$ with $SL(2, \bH)$,
 $Spin(5,1)=SL(2, \bH)$. Then  $Spin(5,1)\cdot Sp(1)$ acts on $\bH^2$ as
  \bea
  (A, a) {\bf v}= A {\bf v} \bar a
  \eea
  where $(A, a)\in Spin(5,1)\cdot Sp(1)$,  ${\bf v}\in \bH^2$ and $A$ acts with a quaternionic matrix multiplication, and where $\bar a$ is the quaternionic conjugate of $a$, $a \bar a=1$.
Using, this it is easy to see
that there is a single non-trivial orbit of $Spin(5,1)\cdot Sp(1)$ on the symplectic Majorana-Weyl spinors
with isotropy group $Sp(1)\cdot Sp(1)\ltimes \bH$. To continue, we have to determine the action of $Sp(1)\cdot Sp(1)\ltimes \bH$
on $\bH^2$. Decomposing $\bH^2=\bR\oplus {\rm Im} \bH\oplus \bH$, where $\bR$ is chosen to be along the first invariant
spinor,  the action of the isotropy group is
\bea
{\rm Im} \bH\oplus \bH\rightarrow a {\rm Im} \bH\bar a \oplus b \bH \bar a~,
\la{act}
\eea
where $(a,b)\in Sp(1)\cdot Sp(1)$.
There are two possibilities. Either the second invariant spinor lies in
${\rm Im} \bH$ or in $\bH$. It cannot lie in both because if there is a non-trivial component in $\bH$, there is
a $\bH$ transformation in $Sp(1)\cdot Sp(1)\ltimes \bH$ such that the component in ${\rm Im} \bH$ can be set to zero.
Now if the second spinor lies in ${\rm Im} \bH$, the isotropy group is $Sp(1)\cdot U(1)\ltimes \bH$. On the other
hand if it lies in $\bH$, the isotropy group is $Sp(1)$. This concludes the analysis for two invariant spinors.

There is no  subgroup in $Spin(5,1)\cdot Sp(1)$ which leaves invariant strictly 3 spinors.  For 4 invariant spinors,
there are two cases to consider. Either all four invariant spinors span
the first copy of $\bH$ in $\bH^2$ and the isotropy group is $Sp(1)\ltimes \bH$, or 2 lie in the first copy
and the other 2  lie in the second copy of $\bH$ in $\bH^2$ and the isotropy group is $U(1)$.
The isotropy group of more than 4 linearly independent spinors is  $\{1\}$.

It remains to find representatives of the solutions to the gravitino KSE  up to  gauge transformations. Observe that the generic holonomy group and the gauge group of the KSEs coincide and both act in the same way on the symplectic Majorana-Weyl spinors.  Repeating the
 analysis we have done to identify the isotropy group of spinors in $Spin(5,1)\cdot Sp(1)$, it is straightforward to find the representatives of the invariant spinors. For example in the case of one invariant spinor, since $Spin(5,1)\cdot Sp(1)$ acts on $\bH^2$ with one non-trivial orbit
  which is dense, any spinor can be chosen as a representative. Moreover the representatives can be expressed
  as forms using the description of spinors as in section \ref{laforms}  The isotropy groups of spinors in $Spin(5,1)\cdot Sp(1)$  as well as representatives
of the invariant spinors have been summarized in table 1.

\begin{table}[ht]
 \begin{center}
\begin{tabular}{|c|c|c|}
\hline
$N$&${\mathrm{Isotropy ~Groups}}$  & ${\mathrm{Spinors}}$ \\
\hline
\hline
$1$  & $Sp(1)\cdot Sp(1)\ltimes \bH$ & $1+e_{1234}$\\
\hline
$2$  & $(Sp(1)\cdot U(1))\ltimes\bH$ & $1+e_{1234}~, ~i(1-e_{1234})$\\
\hline
$4$  & $
Sp(1)\ltimes \bH$ & $1+e_{1234}~, ~i(1-e_{1234})~,~e_{12}-e_{34}~,~i(e_{12}+e_{34})$\\
\hline
\hline
$2$  & $
Sp(1))$ & $1+e_{1234}~, ~e_{15}+e_{2345}$\\
\hline
$4$  & $
U(1)$ & $1+e_{1234}~,~i(1-e_{1234})~, ~e_{15}+e_{2345}~,~ i(e_{15}-e_{2345})$\\
\hline
\end{tabular}
\end{center}
\label{ttt}
\caption{\small
The first column gives the number of invariant spinors, the second column the associated isotropy groups
and the third the representatives of the invariant spinors. Observe that if three spinors are invariant, then there is a fourth one.
Moreover the isotropy group of more than 4 spinors is the identity.}
\end{table}

\vskip 0.5cm

\subsection{Solution of remaining KSEs}

One expects that given some parallel spinors, ie a solution of the gravitino KSE, only some of them will be Killing, ie only some  will also solve the
remaining KSEs. Therefore to find all supersymmetric backgrounds, one has to investigate which of the parallel spinors also solve the remaining KSEs. There are many possibilities  and the analysis
is rather involved. Because of this,  it will not be presented here and can be found in \cite{ap1}.  However the final result is rather straightforward. Apart from one case
that has to do with the hyperini KSE, to identify all supersymmetric backgrounds  suffices to consider the cases where all
parallel spinors also solve the remaining KSEs and so are Killing. The results are summarized in table 2

\begin{table}[ht]
 \begin{center}
\begin{tabular}{|c|c|}\hline
   ${\rm hol}({\cal D})$ &$N$
 \\ \hline \hline
  $Sp(1)\cdot Sp(1)\ltimes\bH$& 1 \\
\hline
$Sp(1)\cdot U(1)\ltimes\bH$&$*$, 2
\\ \hline
$Sp(1)\ltimes\bH$&$*$, $*$, $3$, 4
\\ \hline \hline
$Sp(1)$&$*$, 2
\\ \hline
$U(1)$&$*$, $*$, $-$, 4
\\ \hline
$\{1\}$& $*$,$*$,$*$, $*$,$-$, $-$, $-$, 8
\\ \hline
\end{tabular}
\end{center}
\caption{\small
In the  columns are the holonomy  groups that arise from the solution of the gravitino KSE
and the number $N$  of supersymmetries, respectively. $*$ entries denote the cases that occur
but are special cases of others with the same number of supersymmetries but with less parallel spinors. The $-$  entries denote cases which
do not occur. The Killing spinors for $N=1,2,4$  are the same as those given in table 1 while for $N=3$ the Killing spinors are given in  (\ref{dessp1}).}
\end{table}

 To complete the analysis, it suffices to give the Killing spinors of the $N=3$ case; all the remaining ones
  can be found in table 1. The three Killing spinors can be chosen as
\bea
1+e_{1234}~, ~i(1-e_{1234})~,~~~e_{12}-e_{34}~.
\la{dessp1}
\eea
It turns out that if the gravitino, tensorini and gaugini KSEs admit (\ref{dessp1}) as a solution, then
they admit also $i(e_{12}+e_{34})$ as a solution. Thus all the parallel spinors of this case solve the three out of four KSEs.
However, this is not the case for the hyperini KSE. The conditions that arise on evaluating  the hyperini KSE
on (\ref{dessp1}) are different from those that one finds when the same KSE is evaluated on all 4 $Sp(1)\ltimes \bH$-invariant
spinors. As a result, there is a distinct case preserving strictly 3 supersymmetries.

\newsection{Geometry}

Having found representatives for the Killing spinors, it is straightforward to evaluate the KSEs
and derive the linear systems for all cases. The linear systems can then be solved to derive the
conditions required on the  fields so that the KSEs admit a solution. The analysis is similar to the
paradigm in section \ref{paradigm}. Before, we proceed with a case by case analysis, it is instructive
to first observe that in all cases the solution of the gravitino KSE can be summarized by stating that
the holonomy of the supercovariant connection is included in the isotropy group $G$ of the parallel spinors, ie
\bea
\mathrm{hol}({\cal D})\subseteq G~,
\eea
where all groups $G$ are presented in table 1.
There are several ways that this condition can be expressed in a differential geometric way. One is to consider the
forms constructed as Killing spinor bilinears. Given two spinors $\epsilon_1$ and $\epsilon_2$, one class of
form bilinears is
\bea
\tau={1\over k!} B(\epsilon_1, \gamma_{\mu_1\dots \mu_k} \epsilon_2)\,\, e^{\mu_1}\wedge\dots \wedge e^{\mu_k}~,
\eea
where $B(\epsilon_1, \epsilon_2)=\langle \Gamma_{06789} \epsilon_1^*, \epsilon_2\rangle$ is the Majorana inner product
in the basis chosen in section \ref{laforms}, and where $\langle\cdot, \cdot\rangle$ is the Hermitian inner product on $\Lambda(\bC^5)$. Assuming that $\epsilon_1$ and $\epsilon_2$ satisfy the
gravitino KSE, it is easy to see that
\bea
\hat\nabla_\nu \tau=0~.
\label{par1x}
\eea
The form $\tau$ is covariantly constant with respect to $\hat\nabla$-
the  $Sp(1)$ connection ${\cal C}^{\pr}$ does not contribute in the covariant constancy condition.

Another class of bilinears is the $\mathfrak{sp}(1)$-valued forms
\bea
\tau^{\pr}={1\over k!} B(\epsilon_1, \gamma_{\mu_1\dots \mu_k}\rho^{\pr} \epsilon_2)\,\,
 e^{\mu_1}\wedge\dots \wedge e^{\mu_k}~.
 \eea
 Assuming again that $\epsilon_1$ and $\epsilon_2$ satisfy the gravitino KSE, one finds that
 \bea
 \hat\nabla_\nu \tau^{\pr}+2\, {\cal C}^{\ps}_\nu \epsilon^{\pr}{}_{\ps\pt} \tau^{\pt}=0~.
 \label{par2}
 \eea
Observe that the $\mathfrak{sp}(1)$-valued form bi-linears are twisted with respect to the  $Sp(1)$ connection ${\cal C}^{\pr}$.
So $\tau^{\pr}$ are not forms but rather vector bundle valued forms. However for simplicity in what follows, we shall refer to both
$\tau$ and $\tau^\pr$ as forms.

To solve the gravitino and identify the conditions on the geometry of spacetime, we shall investigate the consequences
of (\ref{par1x}) and (\ref{par2}) in each case. Then we shall  investigate the conditions on the fields imposed
by the remaining KSEs.

\subsection{N=1}

\subsubsection{Gravitino KSE and Spacetime geometry}

To express the form spinor bilinears for backgrounds preserving one supersymmetry, it is convenient
to introduce a lightcone-Hermitian frame on the spacetime, $(e^-, e^+, e^\alpha, e^{\bar\alpha})$, $\alpha=1,2$, ie the metric
is written as
\bea
ds^2=2 e^+ e^-+\delta_{ij} e^i e^j= 2 (e^+ e^-+\delta_{\alpha\bar\beta} e^\alpha e^{\bar\beta})~.
\label{1metr}
\eea
This frame can be chosen such that the form spinor bilinears are
\bea
e^-~,~~~~e^-\wedge \omega^1~,~~~e^-\wedge \omega^2~,~~~e^-\wedge \omega^3~,
\label{n1sb}
\eea
where $e^-$ is a null one-form and
\bea
\omega^1=-i\delta_{\a\bar\b} e^\a\wedge e^{\bar\b}~,~~~\omega^2=-e^1\wedge e^2-e^{\bar 1}\wedge e^{\bar 2}~,~~~
\omega^3=i(e^1\wedge e^2-e^{\bar 1}\wedge e^{\bar 2})~.
\label{3hermf}
\eea
Clearly $\omega^{\pr}$, $\pr=1,2,3$  are Hermitian forms for a quaternionic structure $J_1,J_2,J_3$, $J_\pr J_\ps=-\delta_{\pr\ps} {\bf 1}+ \epsilon_{\pr\ps\pt} J_\pt$,
on the directions transverse to $(e^+, e^-)$.

The conditions that the gravitino KSE imposes on the spacetime geometry can be rewritten as
\bea
\hat\nabla_\mu e^-=0~,~~~\hat\nabla_\mu (e^-\wedge \omega^{\pr})+ 2\,
{\cal C}_\mu^{\ps} \epsilon^{\pr}{}_{\ps \pt} (e^-\wedge\omega^{\pt})=0~.
\la{par1}
\eea
The second equation can be thought as the Lorentzian analogue of the Quaternionic K\"ahler with torsion condition
of \cite{qkt}.
The integrability conditions to these parallel transport equations are
\bea
\hat R_{\mu_1\mu_2, + \nu}=0~,~~~-\hat R_{\mu_1\mu_2,}{}^k{}_i \omega^{\pr}{}_{kj}+(j,i)+2\scrF^{\ps}_{\mu_1\mu_2}
\epsilon^{\pr}{}_{\ps\pt} \omega^{\pt}_{ij}=0~.
\la{1int1}
\eea
In addition to this, the torsion $H$ has to be anti-self-dual in 6 dimensions. The conditions for this in the lightcone-Hermitian frame can be written as
as
\bea
H_{+\a\b}=H_{+\a}{}^{\a}=0~,~~~H_{-+\bar\a}+H_{\bar\a\b}{}^\b=0~,~~~H_{-1\bar 1}-H_{-2\bar2}=0~,~~~H_{-1\bar2}=0~,
\la{self1}
\eea
where $\epsilon_{-+1\bar1 2\bar2}=\epsilon_{013245}=-1$.
Notice that from the 4-dimensional perspective of directions transverse to $(e^+, e^-)$, $H_{+ij}$ is an anti-self-dual while $H_{-ij}$ is a self-dual 2-form,
respectively.

To specify the spacetime geometry, one has to solve (\ref{par1}) subject to (\ref{self1}).
The first condition in (\ref{par1}) implies that
\bea
{\cal L}_X g=0~,~~~de^-=i_X H~.
\la{dexh}
\eea
ie that the  vector field $X$ dual to 1-form $e^-$ is {\it Killing} and the the $i_X H$ component of $H$ is given by the exterior
derivative of the bilinear $e^-$. In fact, $X$ leaves invariant all the fields of the theory.
From this, it is easy to see that the torsion 3-form can be written as
\bea
H=e^+\wedge de^-+{1\over2} H_{-ij} e^-\wedge e^i\wedge e^j+ \tilde H~,~~~\tilde H={1\over3!} \tilde H_{ijk} e^i\wedge e^j\wedge e^k~.
\la{1tor}
\eea
Anti-self-duality of $H$ relates the $\tilde H$ component to $de^-$. In particular, one has that
\bea
\tilde H=-{1\over3!}(de^-)_{-\ell}\,\,\epsilon^\ell{}_{ijk} \,\, e^i\wedge e^j\wedge e^k~.
\la{thde}
\eea

This solves the first condition in (\ref{par1}). To solve the remaining three conditions, consider first the
parallel transport equation in (\ref{par1}) along the light-cone directions. Since $H_{+ij}$ is anti-self-dual, one has that
\bea
{\cal D}_+\omega^{\pr}=\nabla_+\omega^{\pr}+ 2\,
{\cal C}_+^{\ps} \epsilon^{\pr}{}_{\ps \pt} \omega^{\pt}=0~.
\eea
As we shall see from the hyperini KSE (\ref{dpp}), ${\cal C}_+=0$, and so the above condition becomes a restriction on the geometry
\bea
\nabla_+\omega^{\pr}=0~.
\la{1geom1}
\eea
Next
\bea
{\cal D}_-\omega^{\pr}_{ij}=\nabla_-\omega^{\pr}_{ij}-H_-{}^k{}_{[i} \omega^{\pr}_{j]k}+
2\, {\cal C}_-^{\ps} \epsilon^{\pr}{}_{\ps\pt} \omega^{\pt}_{ij}=0~.
\eea
Since $H_{-ij}$ is self-dual, this implies that it can be written as
\bea
H_{-ij}=w_{\pr} \omega^{\pr}_{ij}~,
\eea
for some functions $w_{\pr}$.
Thus
\bea
\nabla_-\omega^{\pr}_{ij}+
 w^{\ps} \epsilon^{\pr}{}_{\ps\pt} \omega^{\pt}_{ij}+
2\, {\cal C}_-^{\ps} \epsilon^{\pr}{}_{\ps\pt} \omega^{\pt}_{ij}=0~.
\la{1geom2}
\eea
This is interpreted as a condition which relates ${\cal C}_-^{\ps}$
to the $H_{-ij}$ components of the torsion. As a result, it can be solved to express $H_{-ij}$ in terms of other fields
and the geometry of spacetime.

To determine the conditions imposed on the geometry from the gravitino KSE in directions transverse
to $(e^+, e^-)$, observe that a generic metric connection in 4 dimensions has holonomy contained in
$Sp(1)\cdot Sp(1)$. Thus the only condition required is  the identification of $Sp(1)$ part of the
$\hat\nabla$ spacetime connection with the $Sp(1)$ part of induced connection from the  Quaternionic K\"ahler
manifold ${\cal Q}$ of the hyper-multiplets. This also follows from the integrability conditions (\ref{1int1}).

Thus to summarize, the spacetime admits a null Killing vector field $X$ whose rotation in the directions
transverse to the light-cone is anti-self-dual, ie
\bea
de^-_{ij}=-{1\over2} \epsilon_{ij}{}^{kl} de^-_{kl}~.
\eea
The geometry is restricted by (\ref{1geom1}). Furthermore,
(\ref{1geom2}) relates the self-dual $H_{-ij}$ component of the torsion to the ${\cal C}_-$ component  of the induced
$Sp(1)$ connection from the Quaternionic K\"ahler manifold of the hyper-multiplets.  The remaining conditions
are given by the integrability conditions (\ref{1int1}). The metric and torsion
of the spacetime can be written as
\bea
ds^2&=&2e^- e^++\delta_{ij} e^i e^j~,
\cr
H&=&e^+\wedge de^-- \big({1\over16}\omega^\pr_{kl} \nabla_-\omega^{\ps kl} \epsilon_{\pr\ps}{}^\pt
+{\cal C}^\pt_-\big )\,\omega_{\pt ij}\,\, e^-\wedge e^i\wedge e^j
\cr
&&~~~~~~~~~~~~~-{1\over 3!} (de^-)_{-\ell}\,\,\epsilon^\ell{}_{ijk} \,\, e^i\wedge e^j\wedge e^k~.
\la{sumn1}
\eea
These are the full set of conditions on the fields and geometry of spacetime for the gravitino KSE  to admit a parallel spinor.

\subsubsection{Gaugini}

Substituting, the Killing spinor $1+e_{1234}$ into the gaugini KSE, one finds that the solution to the linear system is
\bea
F^{\km}_{+i}=F^{\km}_{+-}=0~,~~~F^{\km}_\a{}^\a+i\mu^\km_1=0~,~~~2F^{\km}_{12}+\mu^\km_2-i\mu^\km_3=0~.
\la{1g1}
\eea
As a result, we find that the gauge field can be written
\bea
F^{\km}= F^{\km}_{-i}\, e^-\wedge e^i+{1\over2} \mu^\km_{\pr} \omega^{\pr}+ (F^{\rm asd})^{\km}~,
\eea
where $F^{\km}_{-i}$ and the anti-self-dual components  $F^{\rm asd}_{ij}$  are not restricted by the KSEs. The self-dual part is completely
determined in terms of the moment maps $\mu$.

\subsubsection{Tensorini}

A direct computation of the tensorini KSEs on the spinor $1+e_{1234}$  reveals that
\bea
T^\uM_+=0~,~~~H^\uM_{+\alpha}{}^\alpha=H^\uM_{+\a\b}=0~,
\cr
T^\uM_{\bar\alpha}-{1\over2} H^\uM_{-+\bar\alpha}-{1\over2} H^\uM_{\bar\alpha\beta}{}^\beta=0~.
\la{1tens1}
\eea
Note that the tensorini KSEs commute with the Clifford algebra operations $\rho^{\pr}$ in (\ref{spgen}). As a result, if the tensorini KSE admits a solution $\epsilon$,
then $\rho^{\pr}\epsilon$ also solve the KSE.  As a result, the four spinors
\bea
1+e_{1234}~,~~~\rho^{\pr}(1+e_{1234})~,~~~\pr=1,2,3,
\eea
are  solutions to the tensorini KSE.

The 3-form field strengths are self-dual in 6 dimensions. This implies that
\bea
H^\uM_{-\a\b}=H^\uM_{-\a}{}^\a=0~,~~~H^\uM_{-+\bar\a}-H^\uM_{\bar\a\b}{}^\b=0~,~~~H^\uM_{+1\bar 1}-H^\uM_{+2\bar2}=0~,~~~H^\uM_{+1\bar2}=0~.
\la{aself1}
\eea
Combining these conditions with those from the tensorini KSE, one finds that
\bea
H^\uM_{+ij}=0~.
\eea
Moreover (\ref{aself1}) implies that $H^\uM_{-ij}$ is {\it anti-self-dual} in the directions transverse to $(e^+, e^-)$ and this
component is not otherwise restricted by the KSEs. Therefore, the solution of the KSEs can be expressed as
\bea
T^\uM&=&T^\uM_- e^-+ T^\uM_i e^i~,
\cr
H^\uM&=&{1\over2} H^\uM_{-ij}\, e^-\wedge e^i\wedge e^j+ T^\uM_i e^-\wedge e^+\wedge e^i- {1\over3!}
T^\uM_\ell\,\epsilon^\ell{}_{ijk}\,\,e^i\wedge e^j\wedge e^k ~.
\la{humx}
\eea
In addition,  $T^\uM_i=x^\uM_\ur\partial_i v^\ur$. Substituting this in  (\ref{humx})  all  components of $H^\uM$  apart from $H^\uM_{-ij}$ are expressed in terms of the tensor multiplet scalars.

\subsubsection{Hyperini}
To solve the hyperini KSE, one has to identify the $\epsilon_\uA$ components of the Killing spinor
in the context of spinorial geometry. In our notation $\epsilon^1=1$ and $\epsilon^2=e_{1234}$ and since
$\epsilon_1=-\epsilon^2$ and $\epsilon_2=\Gamma_{34}\epsilon^1$ as in (\ref{hypdef}), one has $\epsilon_1=-e_{1234}$ and $\epsilon_2=e_{34}$. Substituting
these into the KSE, one finds the conditions
\bea
V_+^{\ua\uA}=0~,~~~-V_1^{\ua \uo}+ V_{\bar2}^{\ua \u2}=0~,~~~ V_2^{\ua \uo}+ V_{\bar1}^{\ua \u2}=0~,
\la{1hyp1x}
\eea
where we have set $\ki=\uo, \u2$ to distinguish the range of the $\ki$ index from the range of the holomorphic index
$\alpha=1,2$ of the spacetime.
The conditions (\ref{1hyp1x}) can be expressed in terms of the hyper-multiplet scalars as
\bea
D_+q^\ui=0~,~~~(\tau^i)^\uA{}_\uB D_i q^\ui E^{\uB a}{}_\ui = 0~,
\la{dpp}
\eea
where $(\tau^i)=( -i\sigma_\pr, 1_{2\times 2})$ and $\sigma_\pr$ are the Pauli matrices.
In the gauge $A_+=0$, the fields $q^\ui$ do not dependent on the coordinate $u$  adapted
to the Killing vector field $X=\partial_u$ as expected.
The last condition in (\ref{dpp}) can equivalently be written in a coordinate basis as
\bea
(\mathfrak{I}^i)^\ui{}_\uj D_{i} q^\uj=0~,
\eea
where $(\mathfrak{I}^i)= (I_1, I_2, I_3, 1_{4n\times 4n})$.

\subsection{N=2 non-compact}

There are two cases with $N=2$ supersymmetry distinguished by the isotropy group of
the Killing spinors. If the isotropy group is non-compact $Sp(1)\cdot U(1)\ltimes \bH$, the two Killing spinors
are
\bea
\epsilon_1=1+e_{1234}~,~~~\epsilon_2=i(1-e_{1234})= \rho^1 \epsilon_1~.
\eea
The additional conditions on the fields which arise from the second Killing spinor
can be expressed as the requirement that the KSEs must commute with the Clifford algebra
operation $\rho^1$.

\subsubsection{Gravitino}

It is clear that the gravitino KSE commutes with $\rho^1$, iff
\bea
{\cal C}_\mu^2={\cal C}_\mu^3=0~.
\eea
The form spinor bi-linears are given in (\ref{par1}) and so the full content of gravitino KSE
can be expressed as
\bea
&&\hat\nabla e^-=0~,~~~\hat\nabla (e^-\wedge \omega)=0~,~~~\hat\nabla (e^-\wedge \omega^2)-2\, {\cal C} e^-\wedge\omega^3=0~,
\cr
&&\hat\nabla (e^-\wedge \omega^3)+2\, {\cal C} e^-\wedge\omega^2=0~,
\la{2par}
\eea
where  $\omega=\omega^1$ and ${\cal C}={\cal C}^1$ and $\omega^\pr$ are given in (\ref{3hermf}).

These conditions can be solved as follows.  The first implies the conditions
(\ref{dexh}), ie that the 1-form $e^-$ is associated with a null Killing vector field. The remaining conditions can be solved
 yielding the geometric conditions
 \bea
 &&\nabla_+ \omega_{ij}^\pr=0~,~~~(de^-)_{-\ell}\, \epsilon^\ell{}_{ijk}=(i_{J}\tilde d \omega)_{ijk}~,
 \cr
 &&\nabla_-\omega^2_{ij}-\nabla_-\omega^1_{k[i} (J_3)^k{}_{j]}-{1\over4} \nabla_-\omega^2_{k\ell} \omega^{3k\ell}
\omega^3_{ij}=0~,
\cr
&&\nabla_-\omega^3_{ij}+\nabla_-\omega^1_{k[i} (J_2)^k{}_{j]}+{1\over4} \nabla_-\omega^2_{k\ell} \omega^{3k\ell}
\omega^2_{ij}=0~,
\la{geom12}
\eea
 and the  integrability conditions
 \bea
 &&\hat R_{\mu_1\mu_2,+\nu}=0~,~~~\hat R_{\mu_1\mu_2,ki}\, J^k{}_j-\hat R_{\mu\nu,kj}\, J^k{}_i=0~,~~~
 \cr
 &&-\hat R_{\mu_1\mu_2,ki}\, (J_2)^k{}_{j}+\hat R_{\mu_1\mu_2,kj}(J_2)^k{}_{i}-2\scrF_{\mu_1\mu_2}
 \omega^{3}_{ij}=0~,
 \la{2int}
 \eea
 where we have set $J=J_1$ as this is distinguished from $J_2$ and $J_3$.

 To derive the first condition in (\ref{geom12}) we have used $D_+q^\ui=0$ which follows
 from the hyperini KSE as explained in the $N=1$ case. The second condition in (\ref{geom12}) arises from the solution
 of the second condition in (\ref{par2}). The integrability conditions (\ref{2int}) first restrict the holonomy of the $\hat\nabla$
 connection along the directions transverse to $(e^+, e^-)$ to lie in $U(2)=Sp(1)\cdot U(1)$ and the last condition identifies the
$U(1)$ part of the curvature $\hat R$ with the curvature of ${\cal C}$.

Moreover, one finds the following expressions for some components of the fields
\bea
H_{-ij}=-\nabla_-\omega_{ik}\, I^k{}_j~,~~~{\cal C}_-= {1\over8} \nabla_-\omega^2_{ij} \omega^{3ij}~.
\eea

To  summarize, the gravitino KSE implies that the metric and $H$ can be written as
\bea
ds^2&=&2e^- e^++\delta_{ij} e^i e^j~,
\cr
H&=&e^+\wedge de^--\nabla_-\omega_{ik}\, I^k{}_j\,\, e^-\wedge e^i\wedge e^j
-{1\over 3!} (de^-)_{-\ell}\,\,\epsilon^\ell{}_{ijk} \,\, e^i\wedge e^j\wedge e^k~.
\la{sumn2}
\eea
This concludes the description of the conditions that arise from the gravitino KSE.

\subsubsection{Gaugini}
The gaugini KSE commutes with $\rho^1$, iff
\bea
\mu_2=\mu_3=0~.
\la{2mom}
\eea
As a result, we have that
\bea
F^{\km}= F^{\km}_{-i}\, e^-\wedge e^i+{1\over2} \mu^\km\,  \omega+ (F^{\rm asd})^{\km}~, ~~~\mu^2=\mu^3=0~,
\eea
where $\mu=\mu^1$.

\subsubsection{Tensorini}
A direct substitution of the second Killing spinor into the tensorini KSEs reveals that there are no additional
conditions to those given in (\ref{1tens1}). As we have mentioned the tensorini KSEs commute with all $\rho$
Clifford algebra operations.

\subsubsection{Hyperini}

Combining the restrictions imposed by the second Killing spinor  with those presented in (\ref{1hyp1x})
for the first Killing spinor, one finds
\bea
V_+^{\ua\uA}=0~,~~~V_\a^{\ua \uo}=0~,~~~ V_{\bar\a}^{\ua \u2}=0~,
\la{2hyp1}
\eea
where again $\ki=\uo,\u2$.
These equations can be rewritten as
\bea
D_+q^\ui=0~,~~~ (I_3)^\ui{}_\uj D_i q^\uj = J^j{}_i D_jq^\ui~.
\label{2hyp1xx}
\eea
The last equation is a Cauchy-Riemann type of equation, ie in the absence of gauge fields, $q$'s satisfy a holomorphicity
condition with respect to the pair of complex structures $(J, I_3)$.

\subsection{N=2 compact}
\label{n2horc}
\subsubsection{Gravitino}

The Killing spinors are $\epsilon_1=1+e_{1234}$ and $\epsilon_2=e_{15}+e_{2345}$ as stated in table 1. It is straightforward
to find that a basis in the form spinor bi-linears is given by the  1-forms
\bea
\lambda^a~,~~~a=-,+,{1}~; ~~~e^i~,~~~i=1,2,3~,
\eea
where we have appropriately relabeled the range of the indices $a$ and $i$. Note that the original labeling which  arises from the
 identification of gamma matrices $\gamma$ in (\ref{gammam}) is $a=-, +, 1$ and $i=2,6,7$.

 The conditions implied by the gravitino KSE can be rewritten as
\bea
&&\hat\nabla_\mu \lambda^a=0~,~~~
\cr
&& \hat\nabla_\mu e^i+ 2 \epsilon^i{}_{jk} {\cal C}_\mu^j e^k=0~,
\la{22gravbi}
\eea
where an appropriate identification is chosen between the indices $\pr, \ps$ and $\pt$  which appear in (\ref{par2}) and   $i,j$ and $k$ followed by an appropriate identification of components of ${\cal C}$.

The three 1-forms $\lambda^a$ are parallel with respect to a connection with skew symmetric torsion on the spacetime. As a
result, they are no-where vanishing and their inner product $\eta^{ab}=g(\lambda^a, \lambda^b)$ is constant. In fact, $(\lambda^a, e^i)$ can be used
as a frame on the spacetime and write the metric as
\bea
ds^2=\eta_{ab} \lambda^a \lambda^b+\delta_{ij} e^i e^j~.
\eea
It is clear that the spacetime admits a $3+3$ ``split''. In particular, the tangent space, $TM$, of spacetime decomposes as
\bea
TM=I+\xi~,
\eea
where $I$ is a topologically trivial vector bundle spanned by the  vector fields $X_a$ associated to the three 1-forms $\lambda^a$.

To continue, let us  focus on the first equation in (\ref{22gravbi}). This implies that
\bea
{\cal L}_{X_a} g=0~,~~~d\lambda^a=\eta^{ab} i_b H
\eea
ie $X_a$ are Killing and the $i_b H$ component of $H$ is expressed as the exterior derivative of $\lambda^a$.

 We shall not deal with the most
general case here. This has been done in \cite{ap1}. Instead, we shall assume that the algebra of three
Killing vector fields $X_a$ closes. This together with the anti-self duality of $H$ implies that the only non-vanishing
components of $H$ are $H_{abc}$ and $H_{ijk}$, and
\bea
d\lambda^a={1\over2} H^a{}_{bc} \lambda^b\wedge \lambda^c~,~~~~ H_{abc} \epsilon^{abc}= H_{ijk} \epsilon^{ijk}
\la{geom22}
\eea
for some choice of orientation in $I$ and $\xi$ such that $\epsilon_{abcijk}=\epsilon_{abc} \epsilon_{ijk}$.  The first condition implies that the spacetime
metrically splits locally into a product $G\times \Sigma^3$, where $G$ is a Lorentzian 3-dimensional group
and $\Sigma^3$ is a 3-dimensional manifold. In fact, the Lie algebra of $G$ is
\bea
\bR^{2,1}~,~~~\mathfrak{sl}(2, \bR)~,
\eea
where we have used the classification of Lorentzian Lie algebras in \cite{medina, josec}.

It remains to investigate the geometry of $\Sigma^3$. $\Sigma^3$ is induced with a metric and a 3-form field strength as
\bea
d\tilde s^2(\Sigma^3)=\delta_{ij} e^i e^j~,~~~\tilde H={1\over3!} H_{ijk} e^i\wedge e^j\wedge e^k~,
\eea
which in turn define a connection with skew-symmetric torsion $\hat{\tilde \nabla}$.
Taking the integrability of the second condition
in (\ref{22gravbi}), we find that the curvature of $\hat{\tilde \nabla}$ is
\bea
\hat {\tilde R}_{i_1 i_2, j_1j_2}=-2 \scrF^k_{i_1i_2} \epsilon_{kj_1j_2}~,
\la{int22}
\eea
where we have used ${\cal C}_a=0$ which follows from the hyperini KSE later. This condition implies that the
curvature of $\hat{\tilde \nabla}$  is given in terms of $Sp(1)$ part of the curvature
of the Quaternionic K\"ahler manifold ${\cal Q}$ of the hyper-multiplet scalars  induced on the spacetime.

To summarize, the spacetime is locally a product $G\times \Sigma^3$, where $G$ is a 3-dimensional Lorentzian group and $\Sigma^3$
is a Riemannian manifold, such that
 one has
 \bea
 ds^2=\eta_{ab} \lambda^a \lambda^b+\delta_{ij} e^i e^j~,~~~~H={1\over3!} H_{abc} \lambda^{a}\wedge \lambda^{b}\wedge \lambda^{c}+{1\over3!} H_{ijk}e^i\wedge e^j\wedge e^k~,
 \eea
 provided that the conditions (\ref{geom22}) and (\ref{int22}) hold.

\subsubsection{Gaugini}

Evaluating the gaugini KSE on $e_{15}+e_{2345}$ and combining the resulting conditions  with those of
(\ref{1g1}) that are derived from evaluating the gaugini KSE on the first spinor $1+e_{1234}$, we get that
\bea
F^{\km}=- {1\over2}\epsilon_{ijk} \mu^{\km k}\, e^i\wedge e^j~,
\la{2gau2}
\eea
where again we have appropriately identify the $\pr, \ps, \pt= 1,2,3 $ indices of the moment maps with $i,j,k=1,2,3$, ie with those of the frame on $\Sigma^3$ .
Therefore, the curvature field strengths have support on $\Sigma^3$ and are completely determined
in terms of the moment maps $\mu$.

\subsubsection{Tensorini}

Substituting  $e_{15}+e_{2345}$ into the tensorini KSEs and   comparing the resulting conditions
 with those derived  in (\ref{1tens1}) which arise from evaluating the same KSEs on $1+e_{1234}$, and using the self-duality
of $H^\uM$ (\ref{aself1}), one finds that
\bea
T_\mu^\uM=0~,~~~H^\uM_{\mu\nu\rho}=0~.
\la{thz}
\eea
Expressing of $T$ and $H^\uM$ in terms of the physical fields (\ref{ric}), one finds that the scalars
 are constant and 3-form field strengths of the tensor multiplet vanish.

\subsection{Hyperini}

Evaluating the hyperini KSE on $e_{15}+e_{2345}$ and comparing the results with those of (\ref{1hyp1x}) which arise
from evaluating the same KSE on the first spinor $1+e_{1234}$, we find that
\bea
D_a q^\ui=0~,~~~ D_i q^\ui=-\epsilon_{i}{}^{jk}\, (I_{j})^\ui{}_\uj\, D_{k}q^\uj~.
\la{2hyp2ct}
\eea
In the gauge that $A_a=0$ which can always be chosen locally as $F^\km_{ab}=0$ from the gaugini KSE, one concludes  that $q$
does not dependent on the coordinates of the group $G$.

\newsection{N=4 non-compact}

The four Killing spinors with isotropy group $Sp(1)\ltimes\bH$ of table 1  can be rewritten as
\bea
1+e_{1234}~,~~~\rho^1 (1+e_{1234})~,~~~\rho^2 (1+e_{1234})~,~~~\rho^3 (1+e_{1234})~.~~~
\la{4ncks}
\eea
Therefore for the KSEs to admit these as  Killing spinors they must commute with the Clifford algebra operations $\rho^{\pr}$. This together
with the conditions we have found for backgrounds  to preserve one supersymmetry give the full set of conditions on the
fields in this case.

\subsubsection{Gravitino}

The gravitino KSE commutes with the $\rho^{\pr}$ operations iff ${\cal C}=0$.  The spinor bilinears are in (\ref{n1sb}) but now their conditions read
\bea
\hat\nabla e^-=0~,~~~\hat\nabla (e^-\wedge \omega^{\pr})=0~.
\la{4gravbi}
\eea
Following similar steps to those of the non-compact $N=1$ and $N=2$ cases, the fields can be expressed as
\bea
ds^2&=&2e^- e^++\delta_{ij} e^i e^j~,
\cr
H&=&e^+\wedge de^-- {1\over16}\omega^\pr_{kl} \nabla_-\omega^{\ps kl} \epsilon_{\pr\ps}{}^\pt \,\omega_{\pt ij}\,\, e^-\wedge e^i\wedge e^j
\cr
&&~~~~~~~~~~~~~-{1\over 3!} (de^-)_{-\ell}\,\,\epsilon^\ell{}_{ijk} \,\, e^i\wedge e^j\wedge e^k~.
\la{4sumn1}
\eea
It remains to present the geometric conditions on the spacetime. These are
\bea
\nabla_+\omega^{\pr}&=&0~,~~~de^-_{ij}=-{1\over2} \epsilon_{ij}{}^{kl}\, de^-_{kl}~,
\cr
de^-_{-j}\, \epsilon^j{}_{i_1i_2i_3}&=&(i_{J^{\pr}} \tilde d\omega^{\pr})_{i_1i_2i_3}~,~~~(\mathrm{ no ~\pr ~summation})~.
\la{4geom1}
\eea
To derive these, we have solved (\ref{4gravbi}) and applied the anti-self-duality of $H$.

\subsubsection{Gaugini}
The KSEs commute with $\rho^{\pr}$, iff
\bea
\mu_1=\mu_2=\mu_3=0~.
\la{4mom}
\eea
These are in addition to the conditions given in (\ref{1g1}). Thus, we have that
\bea
F^{\km}= F^{\km}_{-i}\, e^-\wedge e^i+ (F^{\rm asd})^{\km}~.
\eea

\subsubsection{Tensorini}

The tensorini KSE commutes with the Clifford algebra operations $\rho^{\pr}$. Thus
there are no additional
conditions to those given in (\ref{1tens1})

\subsubsection{Hyperini}

The conditions which arise from the hyperini KSEs are
\bea
D_+ q^\tM=D_i q^\tM=0
\eea
Therefore the only non vanishing component of  the  derivative on the scalars is $D_-q^\ui$.

\subsubsection{N=3 descendant}

Unlike all other cases, the $N=4$ backgrounds with $Sp(1)\ltimes \bH$-invariant parallel spinors exhibit an
independent descendant with 3 supersymmetries.  The conditions for this can be easily found by evaluating the
hyperini KSEs on the three spinors  (\ref{dessp1}). The conditions on the scalar $q$ are
\bea
D_+q^\ui=0~,~~~ (J_\pr)^i{}_j D_i q^\ui= (I_\pr)^\ui{}_\uj D_j q^\uj~,
\eea
where $J_r$ are the complex structures transverse to $(e^+, e^-)$ associated with 2-form bilinears $\omega^\pr$ and $I_\pr$ is the quaternionic structure
on the scalar manifold ${\cal Q}$ of the hyper-multiplets. The above condition in the absence of gauge fields implies that $q$'s are locally
quaternionic maps.

\subsection {N=4 compact}
\label{n4horc}

The Killing spinors are  the $U(1)$-invariant spinors of table 1. These can be rewritten as
\bea
1+e_{1234}~,~~~e_{15}+e_{2345}~,~~~\rho^1(1+e_{1234})~,~~~\rho^1(e_{15}+e_{2345})~.
\eea
Thus the conditions on the fields that arise from the KSEs are those we have found for the $Sp(1)$-invariant Killing spinors, and those required for the KSEs to commute with the Clifford algebra operation $\rho^1$.

\subsubsection{Gravitino}

The Clifford algebra operation $\rho^1$ commutes with the gravitino KSE provided that ${\cal C}^2={\cal C}^3=0$.
A basis for algebraically independent spinor bilinears is spanned by the 1-forms
\bea
\lambda^a~,~~~a=-,+,1,\bar 1~,~~~e^i~,~~~i=2,\bar 2~.
\eea
The conditions that arise from gravitino KSE can be rewritten as
\bea
\hat\nabla \lambda^a=0~,~~~\hat\nabla e^i-2\, {\cal C}\, \epsilon^i{}_j e^j=0~,
\la{4grav4}
\eea
where we have set ${\cal C}={\cal C}^1$.

The first condition in (\ref{4grav4}) implies that
\bea
{\cal L}_{X_a} g=0~,~~~i_a H= \eta_{ab} d\lambda^b~,
\label{4cgeom4}
\eea
ie the vector fields $X_a$ associated to $\lambda^a$ are Killing and that the $i_aH$ component of $H$ is given in terms
of the exterior derivative of $\lambda^a$, where $\eta_{ab}=g(X_a, X_b)$ is constant.
It is clear that the spacetime admits a $4+2$ split. In particular, the tangent space $TM=I\oplus \xi$, where
now $I$ is a rank 4 trivial vector bundle spanned by the 4 Killing vectors $X_a$.

To continue, we  assume that the algebra of the four Killing vector field closes, ie $H_{abi}=0$. The more general case
without this assumption  has been presented in \cite{ap1}.  The Lorentzian
4-dimensional Lie algebras have been classified and so the algebra of Killing vector fields $X_a$  must be isomorphic \cite{medina, josec} to one of the following
\bea
\bR^{3,1}~,~~~\mathfrak{sl}(2,\bR)\oplus \mathfrak{u}(1)~,~~~\bR\oplus \mathfrak{su}(2)~,~~~\mathfrak{cw}_4~.
\la{4liealg}
\eea
Furthermore the anti-self duality of $H$ implies that
\bea
H_{aij}={1\over3!}\epsilon_{ij}\,\epsilon_a{}^{b_1b_2b_3} H_{b_1b_2b_3}~,
\eea
where $\epsilon_{abcdij}=\epsilon_{abcd} \epsilon_{ij}$.

Next, we have that
\bea
d\lambda^a-{1\over2} H^a{}_{bc} \lambda^b\wedge \lambda^c={1\over 2} H^a{}_{ij} e^i\wedge e^j~,
\la{4curv4}
\eea
where $H_{abc}$ are the structure constants of the Lie algebra of the four Killing vector fields.
Locally the spacetime can be thought of as a principal bundle with fibre group that has a Lie algebra as
in (\ref{4liealg}),  base space a 2-dimensional manifold $\Sigma^2$ and principal bundle connection $\lambda^a$. In such a case, the rhs of (\ref{4curv4})
is the curvature of $\lambda$ which measures the twist of the fibre over the base space. Since the curvature does not vanish
 the splitting of spacetime is not a product.  This is unlike the $3+3$ splitting of the $N=2$ backgrounds which is a product.
The last condition in (\ref{4grav4}) identifies the spacetime connection along the directions transverse
to the Killing vectors with a $U(1)$ component of the  induced $Sp(1)$ quaternionic K\"ahler  connection. This can also be seen by investigating
the integrability conditions of (\ref{4grav4}). In particular, one finds that the only non-vanishing components
 of the $\hat R$  curvature of spacetime are
\bea
 \hat R_{i_1 i_2, j_1j_2}=-2 \scrF_{i_1i_2}\, \epsilon_{j_1j_2}~.
\la{4curel}
\eea
where we have anticipated the results from the hyperini KSE that ${\cal C}_a=0$.

To summarize, the metric and 3-form field strengths are
\bea
ds^2&=&\eta_{ab} \lambda^a \lambda^b+ \delta_{ij} e^i e^j~,~~
\cr
H&=&{1\over 3!} H_{abc} \lambda^a\wedge \lambda^b \wedge \lambda^c+{1\over2\cdot 3!}\epsilon_{ij}\,\epsilon_a{}^{b_1b_2b_3} H_{b_1b_2b_3} e^a\wedge e^i\wedge e^j~,
\eea
and the geometric conditions are given in (\ref{4cgeom4}) and (\ref{4curel}).

\subsubsection{Gaugini}

The gaugini KSE commutes with $\rho^1$ iff $\mu^2=\mu^3=0$. Combining this with (\ref{2gau2}), one finds
\bea
F^{\km}={1\over2} \mu^{\km} \epsilon_{ij} e^i\wedge e^j~,
\eea
where  $\mu=\mu^1$.

\subsubsection{Tensorini}

The tensorini KSE commutes with all the Clifford algebra $\rho^{\pr}$ operators. Since both $1+e_{1234}$
and $e_{15}+e_{2345}$ are Killng spinors, one concludes that  all 8 supersymmetries are preserved.  Thus
$T^\uM=H^\uM=0$ as in (\ref{thz}). In turn, the tensorini multiplet scalars are constant and the 3-form field strengths
vanish.

\subsubsection{Hyperini}\label{n4compxx}

Evaluating the  hypernini KSEs on the Killing spinors,  one finds
 \bea
 D_a q^\ui=0~,~~~a=-,+,1,\bar 1~,~~~i D_{ 2}q^\ui =(I_3)^\ui{}_\uj D_{ 2}q^\uj.
 \eea
 Clearly, the scalar fields $q$ do not depend on 4 spacetime directions in the gauge $A_a=0$. The last  condition is Cauchy-Riemann
type of equations along the remaining two directions.

\subsection{Trivial isotropy group}
\label{n8horc}

Backgrounds with parallel spinors which have a trivial isotropy group admit 8 parallel spinors. The spacetime
is a Lorentzian Lie group with anti-self-dual structure constants. These have been classified in a similar context in
\cite{jose}. In particular,
 the spacetime is locally isometric to
 \bea
 \bR^{5,1}~,~~~AdS_3\times S^3~,~~~CW_6~,
 \eea
 where the radii of $AdS_3$ and $S^3$ are equal, and the structure constants of $CW_6$ are given by a constant self-dual
  2-form
 on $\bR^4$. Moreover
\bea
\scrF({\cal C})=0~.
\eea
This concludes the conditions which arise from the gravitino KSE.

The gaugini KSEs imply that the gauge field strengths vanish and that $\mu^\pr=0$. The tensorini KSEs imply that
the 3-form field strengths vanish and the tensor multiplet scalars are constants. Similar hyperini KSEs imply that the scalars
$q$ are constant. In turn using (\ref{ric}), the latter gives ${\cal C}=0$.

\newsection{ Black hole horizons}

It is well known that the black hole uniqueness theorems in four dimensions
\cite{israel}-\cite{robinson}
do not extend to five and higher. Specifically in five dimensions, apart
from spherical supersymmetric black holes \cite{bmpv}, there also exist black holes with near horizon topology
$S^1\times S^2$, the black rings \cite{reallbh, ring1}. In more than five dimensions, it is expected that there  are black holes with exotic horizon topologies \cite{gibbons1}-\cite{kunduri}.

The  progress that has made towards understanding the geometry of all solutions to the KSEs of supergravity theories  raises the
possibility that all supersymmetric black hole solutions can be classified. So far this goal has not been attained but some
significant progress has been made towards the classification of all near horizon black hole geometries, see \cite{bhreview} for
a recent review and \cite{gt} for brane horizons. Results in this
direction include the identification of all near horizon geometries of
simple 5- and 6-dimensional supergravities \cite{reallbh, jgdm}. In addition,
all near horizon geometries of  4-dimensional ${\cal N}=1$ supergravity coupled
to any number of vector and scalar multiplets have been
classified  \cite{fourhor} and a similar  result has been established for heterotic horizons \cite{hh}.  The geometries of IIB and 11-dimensional supergravity horizons have been
investigated  in \cite{iibhor, mhor}.  More recently, it has been conjectured that  supersymmetric near horizon black hole geometries exhibit supersymmetry enhancement and are invariant under an $SL(2,\bR)$ symmetry.
The latter property is significant as it illustrates the close relationship between
near horizon geometries and conformal symmetry. The conjecture has been proven for a number of theories
in \cite{index} and has been used to show that there are no asymptotically $AdS_5$ supersymmetric black rings \cite{index, groverrings}. The latter generalizes the result of
\cite{ringkun}
proven under stronger symmetry assumptions.

One of the applications of the solution of the KSEs of  (1,0) supergravity theory coupled to any number of vector, tensor, and scalar multiplets is in the context
of  the near horizon geometries of 6-dimensional black holes which preserve at least one supersymmetry.
In particular, one can show  that 6-dimensional (1,0) supergravity coupled to any number of tensor and scalar multiplets  has two classes of near horizon geometries. One is locally isometric
to $AdS_3\times \Sigma^3$, where $\Sigma^3$ is diffeomorphic to $S^3$, and the other  is locally isometric
to $\bR^{1,1}\times {\cal S}$, where the geometry of ${\cal S}$ depends on the hypermultiplet scalars. These results
have been established in \cite{ap2} and in what follows we shall describe some of the key steps in the proof.

In this review, the main focus is on the $AdS_3\times \Sigma^3$ class. This is because
it exhibits some attractive properties like supersymmetry enhancement and a $\times^2SL(2,\bR)$ invariance which, as it has been mentioned,  are now conjectured  to be
  properties of supersymmetric horizons. These horizons preserve 2, 4 and 8 supersymmetries.
In the latter case, they are locally isometric to $AdS_3\times S^3$ with the radii of the two subspaces
equal.

\subsection{Supersymmetric horizons }

 \subsubsection{Near horizon geometry}

For the application to near horizon geometry of extreme black holes, we shall consider (1,0) supergravity theories
 coupled to any number of tensor and scalar multiplets. The  fields can be written in Gaussian null coordinates \cite{wald}.  Such coordinates always exits
 for extreme, smooth, Killing horizons. In these coordinates, the near horizon fields can be expressed as
\bea
ds^2&=&2 \bbe^+ \bbe^- + \delta_{ij} \bbe^i \bbe^j~,
\cr
G^\ur &=&  \bbe^+ \wedge \bbe^- \wedge \big(d_hS^\ur-N^\ur\big)
- r \bbe^+ \wedge \big(  d_hN^\ur + S^\ur dh \big) + dW^\ur~,
\cr
q^{\underline{I}}&=&q^{\underline{I}}(y)~,~~~~\phi= \phi(y)~,
\la{bhdata}
\eea
where
\bea
\bbe^+ = du~,~~~
\bbe^- = dr + r h+r^2\Delta du ~,~~~
\bbe^i &=& e^i{}_\tP dy^\tP~,
\label{nhbasis}
\eea
 and $d_h S^\ur=dS^\ur-h S^\ur$ and $d_hN^\ur=dN^\ur-h\wedge N^\ur$.
 The spacetime has coordinates $(r, u, y^\tP)$. The black hole horizon section ${\cal S}$ is the co-dimension 2 subspace $r=u=0$ and it is assumed to be {\it compact}, {\it connected}, and {\it without boundary}. The dependence of fields  on light-cone coordinates $(r,u)$ is explicitly given. In addition, $dW^\ur$ are  3-forms,  $h, N^\ur$ are 1-forms, and $S^\ur$ are scalars on the horizon section ${\cal S}$ and  depend only on the coordinates $y$.
$\bbe^i$ is a frame on ${\cal S}$ and depends only on $y$ as well. Both the tensor and  hyper-multiplet scalars depend only on the coordinates of ${\cal S}$.

To find the supersymmetric horizons of 6-dimensional (1,0) supergravity, one has to solve both the field and KSEs of the theory for the fields given in (\ref{bhdata}).
We shall proceed with the solution of KSEs.

\subsubsection{Solution of KSEs}

To continue, we substitute (\ref{bhdata}) into the KSEs (\ref{6kse}) and assume that
the backgrounds preserve at least one supersymmetry.  Furthermore, we identify the stationary Killing vector field $\partial_u$ of the near horizon geometry with the Killing vector constructed as a Killing spinor bilinear.  This may appear as an additional
 restriction but this is not the case as it has been established for the analogous case of heterotic horizons in \cite{index}.
Since the vector Killing spinor bilinear is null, one concludes that
$\Delta=0$.
Moreover,  it turns out that the Killing spinor can always be chosen \cite{ap2}  as
\bea
\e=1+e_{1234}~.
\la{hkspin}
\eea
In such a case, a direct comparison with the expression for the fields for $N=1$ backgrounds in (\ref{sumn1}), (\ref{humx}) and
(\ref{dpp}) implies
that the fields can be rewritten as
\bea
 ds^2&=& 2 \bbe^+ \bbe^-+ \delta_{ij} \bbe^i \bbe^j~,
\cr
H&=&{\bf e}^+ \wedge {\bf e}^- \wedge h+r{\bf e}^+ \wedge dh-{1\over 3!} h_{\ell}\,\,\epsilon^\ell{}_{ijk} \,\, \bbe^i\wedge \bbe^j\wedge \bbe^k~~,
\cr
H^\uM&=& T_i^{\uM}\,\,{\bf e}^- \wedge {\bf e}^+ \wedge {\bf e}^i- {1\over3!}
T^\uM_\ell\,\epsilon^\ell{}_{ijk}\,\,\bbe^i\wedge \bbe^j\wedge \bbe^k ~.
\cr
q^{\ui}&=&q^\ui(y)~,~~~\phi=\phi(y)~,
\la{n1fields}
\eea
where we have used the duality relations of the 3-form field strengths.
 In addition the anti-self duality of $H$ requires that
\bea
dh_{ij}=-{1\over2} \e_{ij}{}^{kl} dh_{kl}~.
\eea
It is clear that $H$ is entirely determined in terms of $h$ while
$H^\uM$ is entirely determined in terms of the scalars $\phi$ of the tensor
multiplets.

After,  rewriting of the fields as in (\ref{n1fields}) and establishing that the Killing spinor is (\ref{hkspin}),
 the gravitino  KSE gives
\bea
\tilde{\cal D}_i (1+e_{1234})=0~,
\la{sgkse}
\eea
where
\bea
\tilde{\cal D}_i=\hat{\tilde \nabla}_i+\mathcal{C}^{r'}_i\rho_{r'}~,
\eea
and $\hat{\tilde \nabla}$ is the connection on ${\cal S}$ with skew-symmetric
torsion $-\star_4 h$. This is just the restriction of the gravitino KSE on ${\cal S}$. One can unveil the geometric content of this equation by
considering the twisted  Hermitian 2-forms $\omega^1, \omega^2,\omega^3$ in  (\ref{3hermf}) constructed as  Killing spinor bi-linears
which are now restricted
on ${\cal S}$. Then,  the integrability condition of (\ref{sgkse}) can be expressed as

\bea
-\hat {\tilde R}_{mn,}{}^k{}_i \omega^{\pr}{}_{kj}+(j,i)+2\scrF^{\ps}_{mn}
\epsilon^{\pr}{}_{\ps\pt} \omega^{\pt}_{ij}=0~,
\la{1int1x}
\eea
where
\bea
\scrF^{\ps}_{mn}=\partial_m q^\ui \partial_n q^\uj \scrF^\ps_{\ui\uj}~.
\eea
This integrability condition identifies the   $Sp(1)\subset Sp(1)\cdot Sp(1)$ component of the
curvature $\hat {\tilde R}$ of the 4-dimensional  manifold  ${\cal S}$ with the pull back with respect to $q$ of the $Sp(1)$ component of the curvature of the Quaternionic K\"ahler manifold ${\cal Q}$.
The restriction imposed on the geometry of ${\cal S}$ by (\ref{1int1x})
depends on the scalars $q^\ui$. In particular, if $q^\ui$ are constant, then
$\scrF_{mn}=0$ and
(\ref{1int1x})  implies that ${\cal S}$ is an HKT manifold \cite{hkt}.

There are no additional conditions arising from the tensorini KSE. The hyperini KSE
requires that $q$ satisfy (\ref{dpp}).
We shall return to the above conditions imposed by the KSEs after  imposing the restrictions on the fields implied
by the field equations of the theory and the compactness of ${\cal S}$.

\subsection{Horizons with $h\not=0$ and holonomy reduction}

\subsubsection{An application of maximum principle}

There are two classes of horizons to consider depending on whether or not $h$ vanishes.
First, we shall consider only the class that $h\not=0$. If $h\not=0$, we  demonstrate that  the number of supersymmetries preserved by the near horizon geometries is always even. For this we shall use the results we have obtained
 from the KSEs for
horizons preserving one supersymmetry and the field equations of the theory.
The methodology we shall  follow to prove this is to compute $\tilde\nabla^2 h^2$ and
apply the maximum principle utilizing the compactness of ${\cal S}$. In particular, one can establish \cite{ap2} that
\bea
\tilde{\nabla}^2 h^2+h^i\tilde{\nabla}_ih^2 = 2\tilde{\nabla}^ih^j\tilde{\nabla}_ih_j  + 4{\partial}_iq^\ui{\partial}_jq^\uj g_{\ui\uj}h^ih^j~,
\label{h4}
\eea
where $\tilde \nabla$ is the Levi-Civita connection of ${\cal S}$ with respect
to $ds^2({\cal S})=\delta_{ij} \bbe^i \bbe^j$ and $\tilde R$ is the associated Ricci tensor.
Applying now the maximum principle using  the compactness of ${\cal S}$, we find
  that $h^2$ is constant and
\bea
\tilde{\nabla}_ih_j&=&0~,~~~h^i{\partial}_iq^\ui=0~.
\la{hphi}
\eea
To establish the latter equation, we have used that the metric $g_{\tM\tN}$  of the  Quaternionic
K\"ahler manifold ${\cal Q}$ is positive definite.
Thus $h$ is a parallel 1-form on ${\cal S}$ with respect to the Levi-Civita connection
and the scalars of the hyper-multiplets are invariant under the action of $h$.

The existence of a parallel 1-form on the horizon section ${\cal S}$ with respect
to the Levi-Civita connection is a strong restriction. First it implies that
the holonomy of $\tilde \nabla$ is contained in $SO(3)\subset SO(4)$,
\bea
{\rm hol} (\tilde \nabla)\subseteq SO(3)~.
\eea
 Moreover
${\cal S}$ metrically  (locally) splits into a product $S^1\times \Sigma^3$, where $\Sigma^3$ is a 3-dimensional manifold. In turn, as we shall see, the near horizon
 geometry is locally a product $AdS_3\times \Sigma^3$. More elegantly the near horizon geometry admits a supersymmetry enhancement from one supersymmetry to two which we explain later.

To prove (\ref{h4}), we first state the field equations  of 6-dimensional supergravity in the absence of vector multiplets as
\bea
R_{\mu \nu}- \frac{1}{4} \varsigma_{\ur\us}{G^\ur_{\mu}}^{\lambda\rho}G^\us_{\nu\lambda\rho} + \partial_{\mu}v^\ur\partial_{\nu}v_\ur
- 2g_{\ui\uj}\partial_{\mu}q^\ui\partial_{\nu}q^\uj&=&0~,\cr
\nabla_{\lambda}\big(\varsigma_{\ur\us}G^{\us\lambda\mu\nu}\big)&=&0~,\cr
\nabla^{\mu}\partial_{\mu}v^\ur + \frac{1}{6}v_\us G^{\us\mu\nu\rho}G^\ur_{\mu\nu\rho}&=&0~,\cr
D_{\mu}\partial^{\mu}q^\ui &=&0~,
\la{feqns}
\eea
where in the last equation it is understood that the Levi-Civita connections
of both the spacetime and the Quaternionic K\"ahler manifold ${\cal Q}$ have been used to covariantize the expression.

Then one finds that
\bea
\tilde{\nabla}^2 h^2 = 2\tilde{\nabla}^ih^j\tilde{\nabla}_ih_j + 2\tilde{\nabla}^i(dh)_{ij}h^j + 2\tilde{R}_{ij}h^ih^j + 2h^j\tilde{\nabla}_j\tilde{\nabla}_ih^i~.
\label{h2}
\eea
The proof of this is given in \cite{hh}.
To proceed, we shall utilize  the field equations to rearrange the above expression
in such a way that we can apply the maximum principle. Using  the Einstein equation
and
\bea
\tilde{R}_{ij}= R_{ij}-\tilde{\nabla}_{(i}h_{j)} + \frac{1}{2}h_ih_j~,
\eea
one finds that
\bea
2\tilde{R}_{ij}h^ih^j&=&-h^2{\partial}_kv_\ur{\partial}^kv^\ur  + 4{\partial}_iq^\ui{\partial}_jq^\uj g_{\ui\uj}h^ih^j - h^i\tilde{\nabla}_ih^2~.
\label{rij}
\eea
The $\mu\nu=+-$  component of the field equation $\nabla_{\lambda}\big(\varsigma_{\ur\us}G^{\us\lambda\mu\nu}\big)$
together with $H^{i+-}=-h^i$ and $H^{{\underline{M}}i+-}=T^{i{\underline{M}}}$ give
\bea
{\partial}_iv_\ur h^i+v_\ur\tilde{\nabla}_ih^i+\tilde{\nabla}_i{\partial}^iv_\ur=0~.
\la{gmn}
\eea
Acting on the above expression with $v^\ur$, we find
\bea
\tilde{\nabla}_ih^i+v^\ur\tilde{\nabla}_i {\partial}^iv_\ur=0~,
\la{xx1}
\eea
where we have used $v_\ur v^\ur=1$.

The field equation of the scalars of the tensor multiplet gives
\bea
v_\ur\tilde{\nabla}_i{\partial}^iv^\ur =0~,
\la{xx2}
\eea
which when combined with (\ref{xx1})  implies that
\bea
\tilde{\nabla}_ih^i=0~.
\la{xx4}
\eea
In addition (\ref{xx2}) and  $v_\ur v^\ur=1$ give
\bea
{\partial}_kv_\ur{\partial}^kv^\ur=0~.
\la{xx3}
\eea
Thus substituting (\ref{rij}) into (\ref{h2}) and using (\ref{xx4}) and (\ref{xx3}),  we find that
\bea
\tilde{\nabla}^2 h^2+h^i\tilde{\nabla}_ih^2 = 2\tilde{\nabla}^ih^j\tilde{\nabla}_ih_j + 2\tilde{\nabla}^i(dh)_{ij}h^j + 4{\partial}_iq^\ui{\partial}_jq^\uj g_{\ui\uj}h^ih^j~.
\label{h3}
\eea
This expression is close to the one required for the maximum principle to apply. It remains to determine $dh$. For this, consider the $jk$-component of the 3-form  field equation to find
\bea
\nabla^i(v_\ur H_{ijk}+x_\ur^{\uM}H^{\uM}_{ijk})=\epsilon_{ijkl} \partial^iv_\ur h^l+
v_\ur\epsilon_{ijkl}\nabla^ih^l =0~,
\eea
which implies that
\bea
dh =0~,
\eea
Substituting this into  (\ref{h3}), we get (\ref{h4}).

\subsubsection{Supersymmetry enhancement}

To demonstrate supersymmetry enhancement for the backgrounds  with $h\not=0$,
let us re-investigate the KSEs for the fields given in (\ref{n1fields}). It is straightforward to see by substituting (\ref{n1fields}) into the KSEs  that
 the general form of a Killing spinor is
\bea
\epsilon=\epsilon_++\epsilon_-=\eta_+-{u\over2} h_i \Gamma^i\Gamma_+ \eta_-+ \eta_-~,~~~\Gamma_\pm\eta_\pm=\Gamma_\pm\epsilon_\pm=0~,
\eea
where $\eta_\pm$ depend only on the coordinates of ${\cal S}$.  In addition
the gravitino KSE requires that
\bea
\hat{\tilde \nabla}_i\epsilon+ \mathcal{C}^{r'}_i\rho_{r'}\epsilon=0~,
\la{grav2}
\eea
the tensorini KSEs implies that
\bea
(1\pm {1\over2}) T_i^\uM\Gamma^i\epsilon_\pm-{1\over12} H^\uM_{ijk} \Gamma^{ijk}\epsilon_\pm=0~,
\eea
and the hyperini KSEs gives
\bea
i\Gamma^i\epsilon_{\pm\kj}V_i^{{\ka\kj}}=0~.
\la{hyper2}
\eea
Next we shall show that both
\bea
\epsilon_1=1+e_{1234}~,~~~\epsilon_2= \Gamma_- h_i \Gamma^i (1+e_{1234})-u k^2 (1+e_{1234})~,
\la{ksp2}
\eea
are Killing spinors, where we have set $k^2=h^2$ for the constant length of $h$. Observe that the second Killing spinor is constructed by setting
$\eta_+=0$ and $\eta_-=\Gamma_- h_i \Gamma^i (1+e_{1234})$.

We have already
solved the KSEs for $\epsilon_1$. Next observe that $\epsilon_2$ solves
the gravitino KSE as the Clifford algebra operation $h_i\Gamma^i \Gamma_-$ commutes
with the supercovariant derivative in (\ref{grav2}) as a consequence of the
reduction of holonomy demonstrated in the previous section. In addition,
the same Clifford operation commutes with the hyperini KSE  as a result of the second eqn in (\ref{hphi}) and (\ref{hyper2}).

It remains to show that $\epsilon_2$ solves the tensorini KSE as well. This
is a consequence of (\ref{xx3}). For this observe that the metric induced on
 $SO(1, n_T)/SO(n_T)$  by the algebraic equation  $\eta_{\ur\us} v^\ur v^\us=1$ is the
 standard hyperbolic metric. So it has definite  signature and as a result,
\bea
\partial_i v^\ur=0~.
\la{xu1}
\eea
Thus, we conclude that the scalar fields are constant and the 3-form
field strengths of the tensorini multiplet vanish.
This  agrees with the classification results of \cite{ap1} for solutions
of the KSEs of 6-dimensional supergravity preserving at least two supersymmetries
whose Killing spinors have   compact isotropy  group and reviewed in section \ref{n2horc}. Some of the results of this section are tabulated in table 3.

\begin{table}
\centering
\fontencoding{OML}\fontfamily{cmm}\fontseries{m}\fontshape{it}\selectfont
\begin{tabular}{|c|c|c|c|}\hline
${\rm Iso}(\eta_+)$& ${\rm hol}({\tilde{\cal D}})$ &$N$&$ \eta_+$
 \\
\hline\hline
  $Sp(1)\cdot Sp(1)\ltimes\bH$ & $Sp(1)$ & $2$ & $1+e_{1234}$
 \\
 \hline
$Sp(1)\cdot U(1)\ltimes\bH$&$U(1)$&$4$&$1+e_{1234}~, ~i(1-e_{1234})$
\\
\hline
$Sp(1)\ltimes \bH^4$&$\{1\}$&$8$&$1+e_{1234}~, ~i(1-e_{1234})~,~e_{12}-e_{34}~,~i(e_{12}+e_{34})$
\\
\hline
\end{tabular}
\label{tab3}
\begin{caption}
{\small {\rm ~~Some of the geometric data used to solving the gravitino KSE are described.
In the first column, we give the isotropy groups, ${\rm Iso}(\eta_+)$, of $\{\eta_+\}$ spinors in $Spin(5,1)\cdot Sp(1)$. In the second column
we state the holonomy of the supercovariant connection $\tilde{{\cal D}}$ of the horizon section ${\cal S}$ in each case. The holonomy of $\hat{\tilde\nabla}$ is identical to that of $\hat\nabla$.
In the third column, we present the number of ${\cal D}$-parallel spinors
and in the last column we give representatives of the $\{\eta_+\}$ spinors.
}}
\end{caption}
\end{table}

\subsection{Geometry}

To investigate the geometry of spacetime, one can compute the
form  bi-linears associated with the Killing spinors (\ref{ksp2}). In particular, one
finds that the spacetime admits 3 $\hat\nabla$-parallel 1-forms given by
\bea
\lambda^- = {\bf e}^-~,~~~\lambda^+ = {\bf e}^+ - \frac{1}{2}k^2u^2{\bf e}^- - uh~,~~~\lambda^1 = k^{-1}(h+k^2u{\bf e}^-)~.
\la{n21forms}
\eea
Moreover, the Lie algebra of the associated vector fields closes in $\mathfrak{sl}(2, \bR)$. To verify this, see \cite{hh}. Since $h$ is $\tilde\nabla$-parallel,  the spacetime is locally metrically a product $
SL(2, \bR)\times \Sigma^3$, ie
\bea
ds^2&=&ds^2(SL(2, \bR))+ ds^2(\Sigma^3)~,
\cr
H&=&d{\rm vol}(SL(2, \bR))+ d{\rm vol}(\Sigma^3)~,
\cr
q^\ui&=&q^\ui(z)~,
\eea
where the scalars of the hyper-multiplet depend only on the coordinates $z$ of
$\Sigma^3$.

In addition to the 1-forms given in (\ref{n21forms}), the spacetime admits
3 more twisted 1-forms bilinears, see \cite{ap1} and section \ref{n2horc}. For the Killing spinors  (\ref{ksp2}), these are given by
\bea
e^{r'}= k^{-1} h_j (J^{r'})^j{}_i \bbe^i~,
\la{ers}
\eea
where $J^{r'}$ is a quaternionic structure on ${\cal S}$  associated with the  twisted Hermitian 2-forms (\ref{3hermf}).

Observe that the frame $e^{r'}$ is orthogonal to $h$ and the rotation between the
$\bbe^i$ and $(h, e^{r'})$ is in $SO(4)$. Therefore $(k^{-1}h, e^{r'})$ is another frame
on ${\cal S}$ with $e^{r'}$ adapted to $\Sigma^3$. Thus
$ds^2({\cal S})=k^{-2} h^2+ds^2(\Sigma^3)$ with  $ds^2(\Sigma^3)=\delta_{r's'}
e^{r'} e^{s'}$.

The metric on $\Sigma^3$ is restricted by  the Einstein equation (\ref{feqns})
and the integrability condition (\ref{1int1x}). The former gives
\bea
R^{(3)}_{r's'}-{1\over2} k^2 \delta_{r's'}-2 \partial_{r'} q^\ui \partial_{s'} q^\uj g_{\ui\uj}=0~,
\la{sigma3}
\eea
where $r', s'$ are indices of $\Sigma^3$ and $R^{(3)}$ is the Ricci tensor of $\Sigma^3$.
This is an equation which determines the metric on $\Sigma^3$ in terms of  $h$
and the hyper-multiplet scalars $q$. The integrability condition (\ref{1int1x})
does not give an independent condition on the metric of $\Sigma^3$.

It remains to find the restriction imposed by supersymmetry on the scalars
$q$ of the hyper-multiplet. Using the results of section \ref{n2horc}, equation (\ref{2hyp2ct}) gives
\bea
\partial_{r'} q^\ui=-\epsilon_{r'}{}^{s't'}\, (I_{s'})^\ui{}_\uj\, \partial_{t'}q^\uj~.
\la{hypereqn2}
\eea
Constant maps are solutions from $\Sigma^3$ into the scalar manifold ${\cal Q}$ of the hyper-multiplet scalars
are solutions.

The geometry on $\Sigma^3$ is determined by (\ref{sigma3}) and depends on the solutions of (\ref{hypereqn2}). For the constant solutions of (\ref{hypereqn2}), $\Sigma^3$
is locally isometric to $S^3$ equipped with the round metric,  and so the near horizon geometry is $AdS_3\times S^3$.

Next suppose  the existence of non-trivial solutions for the equation (\ref{hypereqn2}), and upon substitution the existence of solutions for  (\ref{sigma3}). An priori one expects that the geometry on $\Sigma^3$
 depends on the choice of Quaternionic K\"ahler manifold ${\cal Q}$
 for the hyper-multiplets and the choice of a solution of (\ref{hypereqn2}). However, the differential structure on $\Sigma^3$
 is independent of these choices. To show this first observe that
the Ricci tensor $R^{(3)}$ is strictly positive.  This turns out to be sufficient to determine the topology
 on $\Sigma^3$. To see this note that in 3 dimensions the Ricci
tensor determines the curvature of a manifold. Next, the strict positivity of the
Ricci tensor implies that the (reduced) holonomy of the Levi-Civita connection of $\Sigma^3$ is $SO(3)$. Then a result of Gallot and Meyer, see \cite{ppetersen}, implies that $\Sigma^3$ is a homology 3-sphere. A brief proof of this is as follows. Since
the holonomy of the Levi-Civita connection of $\Sigma^3$ is $SO(3)$, the only parallel
forms are the constant real maps and the volume form of the manifold. On the other
 hand, the positivity of the Riemann curvature tensor implies that all harmonic forms are parallel and the fundamental group is finite. Thus de Rham cohomology of $\Sigma^3$ coincides with that of $S^3$ and so $\Sigma^3$ is a homology 3-sphere. In addition since the fundamental group is finite, the universal cover of $\Sigma^3$ is compact and so by the Poincar\'e conjecture \cite{poincare}  homeomorphic, and so diffeomorphic, to  the 3-sphere.

\newsection{N=4 and N=8 horizons}

\subsection{N=4 horizons}

We have shown that if $h\not=0$, the near horizon geometries preserve 2, 4 or 8
supersymmetries. We have already investigated the case with 2 supersymmetries. The
two additional Killing spinors of horizons with 4 supersymmetries can be chosen as
\bea
\epsilon^3 = i(1-e_{1234})~,~~~\epsilon^4 = -ik^2u(1-e_{1234}) + ih_i\Gamma^{+i}(1-e_{1234})~.
\eea
These horizons are examples of $N=4$ supersymmetric backgrounds with compact isotropy group investigated in section
\ref{n4horc}.
Observe that $\epsilon^3=\rho^1\epsilon^1$ and $\epsilon^4=\rho^1\epsilon^2$. Thus
the KSEs must commute with $\rho^1$.  As a result $\omega_1$ is a well-defined Hermitian form on ${\cal S}$. The 1-form $\hat\nabla$-parallel spinor bilinears are
\bea
\lambda^- &=& {\bf e}^-~,~~~\lambda^+ = {\bf e}^+ - \frac{1}{2}k^2u^2{\bf e}^- - uh~,~~~\lambda^1 = k^{-1}(h+k^2u{\bf e}^-)~,\nn
\lambda^4 &=& e^1~,
\label{n4bilinears}
\eea
where the first 3 bilinears are those of horizons with two supersymmetries and
$e^1$ is given in (\ref{ers}).
The associated vector fields are Killing and their Lie algebra is $\mathfrak{sl}(2, \bR)\oplus \mathfrak{u}(1)$.

The spacetime is locally metrically a product $AdS_3\times \Sigma^3$, as for
horizons preserving 2 supersymmetries. In addition in this case, $\Sigma^3$ is locally
a $S^1$ fibration over a 2-dimensional manifold $\Sigma^2$. The fibre direction
is spanned by $\lambda^4=e^1$. Thus
\bea
ds^2(\Sigma^3)=(e^1)^2+ ds^2(\Sigma^2)~,~~~ds^2({\cal S})=k^{-2} h^2+(e^1)^2+ ds^2(\Sigma^2)~.
\eea
Observe that $de^1\not=0$ as $e^1\wedge de^1$ is proportional to $\tilde H= d{\rm vol}(\Sigma^3)$,  and so the fibration is  twisted.

 It remains
to specify the topology of $\Sigma^2$. For this first observe that from the
results of \cite{ap1} and of section \ref{n4compxx}, the hyper-multiplet scalars depend only on the coordinates
of $\Sigma^2$. Then using  (\ref{sigma3}), one finds that the Ricci tensor of $\Sigma^2$ is positive and so $\Sigma^2$ is a topological sphere. Finally the hyperini KSE implies  that
$q$ are pseudo-holomorphic maps from $\Sigma^2$ into the Quaternionic K\"ahler
manifold ${\cal Q}$.

\subsection{N=8 horizons}

As in the cases with 2 and 4 supersymmetries, one can show that the spacetime
is locally $AdS_3\times \Sigma^3$. In addition for horizons with 8 supersymmetries,
 the  hyperini KSE implies  that the scalars of the hyper-multiplet are constant, see \cite{ap1} and section \ref{n8horc}.  This is compatible with the assertion made in the attractor mechanism, see
\cite{ferrara2} for the 6-dimensional supergravity case, that all the scalars take constant values at the horizon. In such case, the Einstein equation implies that $\Sigma^3$ is locally isometric to $S^3$. Thus the only near horizon geometry preserving 8 supersymmetries with $h\not=0$  is
 $AdS_3 \times S^3$.


\newsection{$(1,0)$-superconformal theories}

As another application, spinorial geometry will be used to investigate the brane solitons  of the KSEs of 6-dimensional
superconformal field theories. A consequence of AdS/CFT correspondence \cite{maldacena} is that the field theory dual of M-theory on the $AdS_7\times S^4$ background is a (2,0) superconformal theory in six dimensions which describes a multiple M5-brane system.
So far  an action  for such a  theory has not been constructed which is local and  6D Lorentz covariant, though there have been suggestions \cite{lp, chu, chu2} which either preserve a subset
of the required symmetries or do not have a general gauge group because of the rigidity in the existence of Euclidean 3-Lie algebras \cite{gp, gg}. In fact it is not apparent that such the (2,0) theory has a classical action as it does not have
 a coupling constant and so a small coupling expansion. Nevertheless if such a theory exists it has to pass  several consistency checks, see eg  \cite{douglas}. These include that after compactification on a circle one should recover the maximally
  supersymmetric gauge theory which describes D5-branes and it should also have a self-dual string and a 3-brane solitons which
  are dictated from the M-brane intersection rules. These  state that a M2-brane ends on a M5-brane on a self-dual string
  and that two M5-branes intersect on a 3-brane \cite{strominger, pktgp}. It is expected   from the perspective
  of a M5-brane theory that the locus of these intersections manifest as worldvolume solitons. The effective dynamics
  of a single M5-brane has been described in \cite{m5h, m5s, m5sh}.

Following a similar strategy to multiple M2-branes \cite{bl, gust} where worldvolume theories were considered preserving less
than maximal supersymmetry \cite{abjm}, the authors of \cite{ssw, ssw3} suggested a class of (1,0) superconformal theories with general gauge groups.
Some of these models admit local actions \cite{ssw, ssw3, bandos} but   suffer from several pathologies which include the non existence of a ground state and possibly the presence of negative norm states. Nevertheless in addition to the classical superconformal
invariance and general gauge group, as we shall show,  exhibit brane solitons in accordance to the M-brane intersection rules, and an intricate mathematical structure \cite{wolf}.

The application of the spinorial geometry to  (1,0) superconformal theories  leads to a systematic solution
of their KSEs and to the construction of explicit self-dual string and 3-brane solitons \cite{ap3, ap4}. The string solutions
are smooth because they are regularized by the size of  instantons.

\subsection{(1,0) superconformal theory and KSEs}

\subsubsection{Fields and KSEs}

The (1,0) superconformal models constructed in \cite{ssw, ssw3} have vector, tensor and hyper-multiplets as well as   appropriate higher form fields which appear  in Stuckelberg-type of couplings.
The field content of the vector multiplets is $(A_\mu^r, \lambda^{\ki r}, Y^{\ki\kj r})$, where $r$ labels the different vector multiplets and $\ki, \kj = 1,2$ are the $Sp(1)$ R-symmetry indices, $A_\mu^r$ are 1-form gauge potentials, $\lambda^{\ki r}$ are symplectic Majorana-Weyl spinors and $Y^{\ki\kj r}$ are auxiliary fields.
The field content of the tensor multiplets is  $(\phi^I, \chi^{\ki I}, B_{\mu\nu}^I)$, where $I$ labels the different tensor multiplets, $\phi^I$ are scalars, $\chi^{\ki I}$ are symplectic Majorana-Weyl spinors, of opposite chirality from those
of the vector multiplets, and $B_{\mu\nu}^I$ are the 2-form gauge potentials.
The field content of the hyper-multiplets are $(q^\tM , \psi^\ka)$, where $q^\tM$ are the hyper-multiplet scalars, which are maps from the spacetime to a  hyper-K\"ahler cone. The latter requires some explanation. Supersymmetry in rigidly supersymmetric theories
requires that the hyper-multiplet scalars take values on a hyper-K\"ahler manifold ${\cal Q}$ instead of a Quaternionic K\"ahler  one
that appears in supergravity. In addition, the existence of superconformal symmetry further restricts the hyper-K\"ahler manifold
to admit a homothetic motion associated with a potential. This is because conformal invariance requires that all  fields
 have a definite scaling dimension. As a result, this makes the hyper-K\"ahler manifold locally a hyper-K\"ahler cone.  $\psi^\ka$ are symplectic Majorana-Weyl spinors of the same chirality as $\chi^{\ki I}$.

The field strengths of the 1- and 2-form gauge potentials associated with the vector and tensor multiplets are
\be
\mathcal{F}_{\mu\nu}^r &\equiv& 2\partial_{[\mu}A_{\nu]}^r - f_{st}{}^rA_\mu^s A_\nu^t + h_I^rB_{\mu\nu}^I~,\\
\mathcal{H}_{\mu\nu\rho}^I &\equiv& 3D_{[\mu}B_{\nu\rho]}^I + 6d_{rs}^I A_{[\mu}^r\partial_\nu A_{\rho]}^s - 2f_{pq}{}^s d_{rs}^I A_{[\mu}^r A_\nu^p A_{\rho]}^q + g^{Ir}C_{\mu\nu\rho r}~,
\ee
respectively, where $f_{rs}{}^t$   $h_I^r, g^{Ir}$ and $d_{rs}^I=d_{(rs)}^I$ are coupling constants, and $C_{\mu\nu\rho r}$ are  three-form gauge potentials introduced
via a St\"uckelberg-type of coupling. In addition,
\be
D_\mu \Lambda^s \equiv \partial_\mu \Lambda^s+ A_\mu^r(X_r)_t{}^s \Lambda^t~,~~~~D_\mu \Lambda^I \equiv \partial_\mu \Lambda^I+ A_\mu^r(X_r)_J{}^I \Lambda^J~,
\ee
where $X_r$ are given by
\bea
 (X_r)_t{}^s=-f_{rt}{}^s+d^I_{rt} h^s_I~,~~~(X_r)_J{}^I=2 h^s_J d^I_{rs}-g^{Is} b_{Jsr}~.
 \eea
The various coupling satisfy a long list
 \bea
 2(d^J_{r(u} d^I_{v)s}-d^I_{rs} d^J_{uv}) h^s{}_J&=& 2 f_{r(u}{}^s d^I_{v)s}-b_{Jsr} d^J_{uv} g^{Is}~,
 \cr
 (d^J_{rs} b_{Iut}+ d^J_{rt} b_{Isu}+2 d^K_{ru} b_{Kst} \delta^J_I) h^u_J&=& f_{rs}{}^u b_{Iut}+f_{rt}{}^u b_{Isu}+ g^{Ju} b_{Iur}
 b_{Jst}~,
 \cr
 f_{[pq}{}^u f_{r]u}{}^s-{1\over3} h_I^s d^I_{u[p} f_{qr]}{}^u&=&0~,
 \cr
 h_I^r g^{Is}&=&0~,
 \cr
 f_{rs}{}^t h^r_I-d^J_{rs} h^t_J h^r_I&=&0~,
 \cr
 g^{Js} h_K^r b_{Isr}-2 h_I^s h_K^r d^J_{rs}&=&0~,
 \cr
 -f_{rt}{}^s g^{It} + d^J_{rt} h^s_J g^{It}- g^{It} g^{Js} b_{Jtr}&=&0~.
 \la{concon}
 \eea
of restrictions required by gauge invariance  established in \cite{ssw}.
In addition, these models are described by an action provided there is a maximally split signature metric\footnote{Since the metric is maximally split,
the kinetic energy of some of the fields is negative which may lead to ghosts in the spectrum. This is an issue affecting this class of theories.}  $\eta_{IJ}$ such that
\bea
g^{Ir}=\eta^{IJ} h^r_I~,~~~d^I_{rt}={1\over2}\eta^{IJ}b_{Jrt}~.
\eea
From now on, the indices $I,J$ are raised and lowered with $\eta$.

To couple hyper-multiplets to the above system \cite{ssw3}, one assumes that the hyper-K\"ahler cone ${\cal Q}$  admits tri-holomorphic isometries generated by the
vector fields $X_{(\mathfrak{m})}=X^\tM_{(\mathfrak{m})}\partial_\tM$, ie isometries which leave also the three
complex structures of the hyper-K\"ahler space invariant. Typically only some of the vector multiplets will be gauged. For this, introduce the embedding tensor $\theta^\mathfrak{m}_r$
and define
\be
A^{\mathfrak{m}} = A^r \theta_r{}^{\mathfrak{m}}~,~~~ \lambda^{\mathfrak{m}} = \lambda^r\theta_r{}^{\mathfrak{m}}~,~~~Y^{\mathfrak{m}}_{\ki\kj} = Y^r_{\ki\kj} \theta_r{}^{\mathfrak{m}}~,
\label{theta1}
\ee
where for consistency with the gauge transformations
\be
h^r{}_I\theta_r{}^{\mathfrak{m}} = 0~,~~~~~~f_{rs}{}^t\theta_t{}^{\mathfrak{m}} = \theta_r{}^{\mathfrak{n}}\theta_s{}^{\mathfrak{p}}f_{\mathfrak{n}\mathfrak{p}}{}^{\mathfrak{m}}~,
\label{theta2}
\ee
and where $[X_{(\mathfrak{n})}, X_{(\mathfrak{p})}]=-f_{\mathfrak{n}\mathfrak{p}}{}^{\mathfrak{m}} X_{(\mathfrak{m})}$.
The KSEs of the model, which are the vanishing conditions for the supersymmetry transformations of the fermions evaluated at the locus where all fermions vanish, are
\bea
\delta \lambda^{ir} &=& \frac{1}{8}\mathcal{F}_{\mu\nu}^r\gamma^{\mu\nu} \epsilon^\ki - \frac{1}{2}Y^{\ki\kj r}\epsilon_\kj + \frac{1}{4}h_I^r\phi^I\epsilon^\ki=0~,\cr
\delta \chi^{\ki I} &=& \frac{1}{48}\mathcal{H}_{\mu\nu\rho}^I \gamma^{\mu\nu\rho}\epsilon^\ki + \frac{1}{4}D_\mu\phi^I \gamma^\mu \epsilon^\ki =0~,\cr
\delta \psi^\ka &=& \frac{1}{2}D_\mu q^\tM \gamma^\mu \epsilon_\ki E^{\ki\ka}{}_{\tM} =0~,
\label{ckse}
\eea
where
\be
D_\mu q^\tM = \partial_\mu q^\tM -A^{\mathfrak{m}}_\mu X^\tM_{(\mathfrak{m})}~.
\ee
In addition, $E^\tM_{\ki\ka}$ is the symplectic frame of the hyper-K\"ahler cone, ie the hyper-K\"ahler metric and hypercomplex structure   are given as
\bea
g_{\tM\tN}= \epsilon_{\ki\kj} \epsilon_{\ka\kb} E^{\ki\ka}{}_{\tM} E^{\kj\kb}{}_{\tN}~,~~~(I_\tau)^\tM{}_\tN=-i\, (\sigma_\tau)^\ki{}_\kj\, \delta^\ka{}_\kb\, E_{\ki\ka}^\tM\, E^{\kj\kb}_\tN~,~~~
\label{qmcs}
\eea
where $\epsilon_{\ki\kj}$ and $\epsilon_{\ka\kb}$ are the symplectic (fundamental) forms of $Sp(1)$ and $Sp(n)$, respectively,  and $\sigma_\tau,  \tau=1,2,3$ are the Pauli matrices. In analogy with similar variations in 6-dimensional (1,0) supergravity, we refer to these KSEs as the gaugini, tensorini and hyperini KSEs, respectively.

The Lagrangian for these theories consist of two parts. One  part, $\mathcal{L}_{VT}$,  involves the vector and tensor multiplets, and the second part, $\mathcal{L}_H$, contains the hyper-multiplets. These two parts are independently supersymmetric and the supersymmetry transformation of the vector multiplets used in the coupling of the hyper-multiplets in $\mathcal{L}_H$ is obtained by contraction with the embedding tensor.

\subsubsection{Field equations }

The field equations of the system are
\bea
D^\mu D_\mu \phi^I &=& -\frac{1}{2}d_{rs}^I(\mathcal{F}_{\mu\nu}^r\mathcal{F}^{\mu\nu s} - 4Y_{ij}^rY^{ijs}) - 3d_{rs}^Ih_J^rh_K^s\phi^J\phi^K~,\label{eqa}
\cr
b_{Irs}Y_{\ki\kj}^s\phi^I &=&{1\over2\lambda} \theta_r{}^\km\, \mu_{\km \ki\kj} ~,\label{eqb}
\cr
b_{Irs}\mathcal{F}_{\mu\nu}^s\phi^I &=&\frac{1}{4!}\epsilon_{\mu\nu\lambda\rho\sigma\tau}\mathcal{H}_r^{(4)\lambda\rho\sigma\tau}~,
\cr
g_{\tM\tN} \nabla_\mu D^\mu q^\tN&=&-Y^\km_{\ki\kj} \partial_\tM \mu^{\ki\kj}_{(\km)}~,
\label{eqc}
\eea
where
\bea
\nabla_\mu D^\mu q^\tM=\partial_\mu D^\mu q^\tM+\Gamma^\tM_{\tN\tP} D^\mu q^\tN  D_\mu q^\tP-\partial_\tN X_{(\km)}^\tM  \theta^\km_r A_\mu^r D^\mu q^\tN~,
\eea
$\lambda$ is a constant, and $\mu_{(\mathfrak{m})\tau}$,
\bea
X_{(\mathfrak{m})}^\tN (\omega_\tau)_{\tN\tM}=-\partial_\tM\mu_{(\mathfrak{m})\tau}~,~~~(\omega_\tau)_{\tM\tN}=g_{\tM\tP}(I_\tau)^\tP{}_\tN~,
\la{moment}
\eea
are the moment maps.
Observe that generically the theory has a cubic scalar field interaction and so the potential term is not bounded
from below. These field equations are also supplemented with
the Bianchi identities
\bea
D_{[\mu}\mathcal{F}_{\nu\rho]}^r &=& \frac{1}{3}h_I^r\mathcal{H}_{\mu\nu\rho}^I~,
\cr
D_{[\mu}\mathcal{H}_{\nu\rho\sigma]}^I &=& \frac{3}{2}d_{rs}^I\mathcal{F}_{[\mu\nu}^r\mathcal{F}_{\rho\sigma]}^s + \frac{1}{4}g^{Ir}\mathcal{H}_{\mu\nu\rho\sigma r}^{(4)}~,
\cr
D_{[\mu}\mathcal{H}^{(4)}_{\nu\lambda\rho\sigma]r}&=&-4d_{Irs} {\mathcal F}^s_{[\mu\nu} {\mathcal H}^I_{\lambda\rho\sigma]}+{1\over5}\theta_r{}^\km {\cal H}^{(5)}_{\km \mu\nu\lambda\rho\sigma}~,
 \label{idh}
\eea
where $\mathcal{H}_{\mu\nu\rho\sigma r}^{(4)}$ is the field strength of the 3-form, and the duality relations
\bea
{1\over 5!} \epsilon_{\mu\nu\rho\lambda\sigma \tau} \theta_r{}^\km \mathcal {H}^{(5)\nu\rho\lambda\sigma \tau}_\km =(X_r)_{IJ} \phi^I D_\mu \phi^J+{2\over\lambda} \theta_r{}^\km X_{(\km)}{}_\tM D_\mu q^\tM~,
\eea
ie the 5-form field strength is dual to the hyper-multiplet scalars.

\subsubsection{KSEs revisited}

The KSEs of the system are the vanishing conditions of the supersymmetry variantions of the fermions given in (\ref{ckse}).
These KSEs are very similar to the (1,0) supergravity KSEs. The only differences are that there is no gravitino KSE
and there is some relabeling of the fields, ie there are three instead of four KSEs the gaugini, tensorini and hyperini ones. Because of this, they can be rewritten in a basis where the
symplectic Majorana-Weyl spinors are identified with the $SU(2)$ Majorana-Weyl spinors of $Spin(9,1)$ as in (\ref{smw}).  In particular, the KSEs can now be rewritten as
\be
\frac{1}{4}\mathcal{F}_{\mu\nu}^r\gamma^{\mu\nu}\epsilon + (Y^{r})_\pr\rho^\pr \epsilon + \frac{1}{2}h_I^r\phi^I\epsilon &=& 0~,\label{ksev12}\\
\frac{1}{12}\mathcal{H}^I_{\mu\nu\rho}\gamma^{\mu\nu\rho}\epsilon + D_\mu\phi^I\gamma^\mu\epsilon &=& 0~, \label{ksev22}
\\
 \frac{1}{2}D_\mu q^\tM \gamma^\mu \epsilon_\ki E^{\ki\ka}{}_{\tM} =0~, \label{ksev23}
\ee
where we have set
\be
-Y^{\ki\kj r}\epsilon_\kj = (Y^{r})_\pr \rho^\pr \epsilon^\ki~,
\ee
and it is understood that
\bea
\epsilon_1 = -\epsilon^2~, ~~~\epsilon_2 = \Gamma_{34}\epsilon^1~,
\eea
and where $\rho$'s are given in (\ref{spgen}).  The latter identification applies in the context of hyperini KSE.

\subsubsection{Solution of KSEs}

To solve the KSEs, it is essential to note that the spinorial geometry method is not sensitive to the way that
the components of the KSEs in the Clifford algebra expansion  depend on the physical fields. Since the KSEs
 of the (1,0) superconformal theory (\ref{ksev12}), (\ref{ksev22}) and (\ref{ksev23})  have the same lexicographic structure as
 the gaugini, tensorini and hyperini KSEs of (1,0) supergravity (\ref{kkk}), the method developed to solve the latter also
 applies to solve the former.  In fact, the analysis is simpler than that of the supergravity theory as one does not have to solve the gravitino KSE.
 The results are summarized in two tables. In table 4, the isotropy groups of the Killing spinors in $Spin(5,1)\cdot Sp(1)$ are given and a choice of
 representatives for the invariant spinors, while in table 5 the number of supersymmetries preserved in each case is denoted.
 Note that for the hyperini KSE there is a distinct case preserving 3 Killing spinors. The Killing spinors can be chosen as in (\ref{dessp1}).


\begin{table}[ht]
 \begin{center}
\begin{tabular}{|c|c|c|}
\hline
$N$&${\mathrm{Isotropy ~Groups}}$  & ${\mathrm{Invariant ~Spinors}}$ \\
\hline
\hline
$1$  & $Sp(1)\cdot Sp(1)\ltimes \bH$ & $1+e_{1234}$\\
\hline
$2$  & $(Sp(1)\cdot U(1))\ltimes\bH$ & $1+e_{1234}~, ~i(1-e_{1234})$\\
\hline
$4$  & $
Sp(1)\ltimes \bH$ & $1+e_{1234}~, ~i(1-e_{1234})~,~e_{12}-e_{34}~,~i(e_{12}+e_{34})$\\
\hline
\hline
$2$  & $
Sp(1))$ & $1+e_{1234}~, ~e_{15}+e_{2345}$\\
\hline
$4$  & $
U(1)$ & $1+e_{1234}~,~i(1-e_{1234})~, ~e_{15}+e_{2345}~,~ i(e_{15}-e_{2345})$\\
\hline
\end{tabular}
\end{center}
\label{ttt1}
\caption{\small
The first column gives the number of invariant spinors, the second column the associated isotropy groups
and the third column representatives of the invariant spinors. Observe that if 3 spinors are invariant, then there is a fourth one
which is also invariant under the same isotropy group.
Moreover the isotropy group of more than 4 linearly independent spinors is the identity.}
\end{table}
{}
\vskip 1.0cm
{}

\begin{table}[ht]
 \begin{center}
\begin{tabular}{|c|c|c|c|}\hline
   ${\rm Isotropy ~Groups}$ &${\rm Gaugini} $& ${\rm Tensorini}$&${\rm Hyperini}$
 \\ \hline \hline
  $Sp(1)\cdot Sp(1)\ltimes\bH$& 1 &$4$& $1$\\
\hline
$Sp(1)\cdot U(1)\ltimes\bH$& 2 &$4$&$2$
\\ \hline
$Sp(1)\ltimes\bH$& 4&$4$&$3,4$
\\ \hline \hline
$Sp(1)$& 2&$8$&$2$
\\ \hline
$U(1)$& 4&$8$&$4$
\\ \hline
$\{1\}$&  8&$8$&$8$
\\ \hline
\end{tabular}
\end{center}
\caption{\small
In the first column the isotropy  groups of the Killing spinors of the gaugini KSE are given. In the second, third  and fourth columns
  the number of Killing spinors of the gaugini, tensorini and hyperini KSEs are stated, respectively. The isotropy groups of the Killing spinors of the tensorini KSE are either $Sp(1)\ltimes \bH$ or $\{1\}$.    The cases that do not appear in the table do not independently  occur.}
\end{table}

Having identified the Killing spinors and the fractions of supersymmetry preserved, it is straightforward to derive the
linear system in each case and solve it to find  the conditions on the fields required by supersymmetry. Since the
spacetime is flat, the task is rather straightforward and it follows closely the analysis we have already presented
for supergravity.  So instead of repeating the details, only the final result will be stated in each case with a minimum
explanation.

\subsubsection{N=1 solutions}
As in the case of supergravity, the KSEs can be easily expressed after choosing a light-cone Hermitian
coordinate system for the 6-dimensional Minkowski spacetime metric. In particular, one writes
\bea
ds^2= 2 e^- e^++ \delta_{ij} e^i e^j=e^- e^++2 \delta_{\alpha\bar\beta} e^\alpha e^{\bar \beta}= 2 dx^+ dx^-+2
 \delta_{\alpha\bar\beta} dz^\alpha dz^{\bar\beta}~,
 \eea
and  assumes the apparent identification between the frame $(e^+, e^-, e^\alpha, e^{\bar\alpha})$ which appears in
 supergravity and the coordinates $(x^+, x^-, z^\alpha, z^{\bar\alpha})$ of  the Minkowski spacetime.

The solution of the gaugini KSEs (\ref{ksev12}) can be expressed as
\bea
\mathcal{ F}^r&=&-h_I^r\phi^I \, e^-\wedge e^++\mathcal{ F}^r_{-i}\, e^-\wedge e^i+ (Y^{r})_{s'} \omega^{s'}+\mathcal{ F}^{{\rm asd},r}~,~~~
\la{fum}
\eea
where $\omega_{s'}$ are the twisted Hermitian forms in (\ref{3hermf}).

Similarly, the tensorini KSEs (\ref{ksev22}) give
\bea
\mathcal{H}^I&=&{1\over2} \mathcal{H}^I_{-ij}\, e^-\wedge e^i\wedge e^j- D_i\phi^I e^-\wedge e^+\wedge e^i+{1\over3!}
D_\ell\phi^I\,\epsilon^\ell{}_{ijk}\,\,e^i\wedge e^j\wedge e^k ~,~~~
\cr
D_+\phi^I&=&0~,
\la{hum}
\eea
where  $\mathcal{H}^I_{-ij}$ is {\it anti-self-dual} in the directions transverse to $(e^+, e^-)$.
Unlike the gaugini KSEs, the tensorini KSEs exhibit supersymmetry enhancement. In particular, if they admit
one Killing spinor $\epsilon$, they also admit three additional Killing spinors given by $\rho^1\e, \rho^2\e$ and $\rho^3\e$. For $\e=1+e_{1234}$, all four Killing spinors are given by the $Sp(1)\ltimes \bH$ invariant  spinors
of table 4.

Next the hyperini KSE gives that
\bea
D_+q^\tM = 0~,~~~(\mathfrak{I}^i)^\tM{}_\tN D_{i} q^\tN=0~,
\eea
where $(\mathfrak{I}^i)= (I_\tau, 1_{4n\times 4n})$ and $I_\tau$ have been given in (\ref{qmcs}).

\subsubsection{$N=2$ solutions with non-compact isotropy group}

The solution to the gaugini KSEs can be expressed as
 \bea
\mathcal{ F}^r=-h_I^r\phi^I \, e^-\wedge e^++\mathcal{ F}^r_{-i}\, e^-\wedge e^i+ Y^r \omega+\mathcal{ F}^{{\rm asd},r}~,
\la{fum2n}
\eea
 where we have set $Y^r=(Y^r)_1$. In this case,  $(Y^{r})_2=(Y^r)_3 =0$. As we have explained in the
 previous section, the tensorini KSEs give the same conditions as in the $N=1$ case.

 The conditions imposed by the hyperini KSEs on the fields can be expressed as
\be
D_+q^\tM = 0~, ~~~D_i q^\tN (I_3)^{\tM}{}_{\tN} = J^{j}{}_{i} D_{j}q^\tM~,
\ee
where $I_3$ is defined in (\ref{qmcs}) and $J^i{}_j = (i\delta^{\alpha}{}_{\beta}, -i\delta^{\bar{\alpha}}{}_{\bar{\beta}})$.
  In the absence of gauge
fields, the above condition becomes the Cauchy-Riemann  equation and $q$ is a holomorphic map from the transverse space to the $(e^+, e^-)$ to the hyper-K\"ahler cone ${\cal Q}$ with respect to the indicated pair of complex structures. The choice  complex structures
 depends on the choice of representatives for the Killing spinors.

\subsubsection{$N=2$ solutions with compact isotropy group }

From the analysis of the supergravity KSEs, we know that the spacetime admits a 3+3 split. This split can be expressed by  splitting the spoacetime index as $\mu=(a, i)$,
where $a=+,-,1$ and $i$ labels the remaining three coordinates, ie the metric is written as
\bea
ds^2= \eta_{ab} e^a e^b+ \delta_{ij} e^i e^j=\eta_{ab} dx^a dx^b+ \delta_{ij} dx^i dx^j~.
\eea
In this notation, the gaugino KSEs give
\bea
\mathcal{F}^r=-\varepsilon_{ijk}(Y^r)^k\, e^i\wedge e^j~,~~~h_I^r\phi^I=0~,
\la{fumn2c}
\eea
where we have appropriately identified the spacetime index with that which labels the auxiliary fields $Y$.

The tensorini KSEs imply that
\be
\mathcal{H}_{\mu\nu\rho}^{I} = 0~,~~~~~~~~~D_\mu\phi^I = 0~. \label{compact2conds}
\ee
Clearly in this case, the tensorini KSEs preserve all 8 supersymmetries. Moreover, the integrability of the last condition
in (\ref{compact2conds}) implies that
\bea
F^r_{\mu\nu} X_{rJ}{}^I \phi^J=0~,
\eea
where $F^r_{\mu\nu}=2\partial_{[\mu}A_{\nu]}^r + X_{st}{}^r A_\mu^s A_\nu^t$.

Finally, the hyperini KSEs give
\be
D_a q^\tM =0~,~~~D_{i}q^\tM = -\epsilon_{i}{}^{jk}(I_{j})^\tM{}_\tN D_{k}q^\tN~,~~~
\label{cn2}
\ee
as in section \ref{n2horc}.

\subsubsection{N=4 solutions with non-compact isotropy group }

The gaugini KSEs give
\bea
\mathcal{ F}^r=-h_I^r\phi^I \, e^-\wedge e^++\mathcal{ F}^r_{-i}\, e^-\wedge e^i+ \mathcal{ F}^{{\rm asd},r}~,
\la{fn4nc}
\eea
where now $Y^1=Y^2=Y^3=0$. The tensorini KSE gives the same conditions as those in the $N=1$ case.
It remains to solve the hyperini KSE. This gives that the only non-vanishing component is
$D_-q^\tM$.

\subsubsection{N=3 Non-Compact}
The hyperini KSE admits a special case that preserves 3 supersymmetries.  The conditions for this are
\bea
D_+q^\tM  = 0~,~~~(J_\tau)^j{}_i D_j q^\tM=(I_\tau)^\tM{}_\tN D_i q^\tN~,~~~\tau=1,2,3~,
\la{quatr}
\eea
for an appropriate choice of a hypercomplex structure $J_\tau$ in the directions transverse to $(e^+, e^-)$. Therefore in the absence of gauge couplings, the hyper-scalars are quaternionic maps. Clearly, the directions transverse to $(e^+, e^-)$ can be identified with the quaternions $\bH$. If the Obata curvature of the the hyper-K\"ahler cone vanishes, then it is possible to introduce quaternionic coordinates on the hyper-K\"ahler cone. In such a case $q$'s can be written as  quaternions ${\bf q}$ and (\ref{quatr}) implies that ${\bf q}={\bf q}({\bf x}, x^-)$, ${\bf x}\in \bH$.

\subsubsection{N=4 solutions with compact isotropy group}

The spacetime admits a 4+2 split. The metric can be written as
\bea
ds^2=\eta_{ab} e^a e^b+ \delta_{ij} e^i e^j= \eta_{ab} dx^a dx^b+ 2 dz^2 dz^{\bar 2}~,
\eea
ie the spacetime index $\mu=(a, i)= (a, 2, \bar 2)$
The tensorini KSEs give that
 \bea
 \mathcal{H}_{\mu\nu\rho}=D_\mu\phi=0~.
 \eea

 The gaugini KSEs imply
 \be
\mathcal{F}^r=-2i Y^r e^2\wedge e^{\bar2}~,~~~~~~h_I^r\phi^I=0~,~~~~~~
\la{4ksecon}
\ee
where we have set $Y^r=(Y^r)_1$.

Next  the hyperini KSEs give
\bea
D_a q^\tM = 0~,~~~J^{j}{}_{i} D_{j} q^\tM =  (I_3)^{\tM}{}_{\tN}  D_{i} q^\tN~,~~~
\la{n4compb}
\eea
where $J^{i}{}_{j}=(i \delta^2{}_2, -i \delta^{\bar 2}{}_{\bar 2})$.

\subsubsection{Maximally supersymmetric solutions}
As we have mentioned all backgrounds which preserve more than 4 supersymmetries are maximally supersymmetric.
It is straightforward to see that the conditions on the fluxes for maximally supersymmetric backgrounds are
\bea
D_\mu\phi^I=0~,~~~h_I^r\phi^I = 0~,~~~\mathcal{F}_{\mu\nu}^r = 0~,~~~\mathcal{H}_{\mu\nu\rho}^{I} = 0~,~~~Y^{\ki\kj r} = 0~, ~~~D_\mu q^\tM=0~.
\eea
Thus all the scalars $\phi^I$ and $q^\tM$ are  covariantly constant. In addition, those projected   by $h$ are required to vanish. Similarly the 2-form and 3-form
field strengths vanish as well. The same applies for the auxiliary fields $Y$.


\subsection{Self-dual string solitons}

\subsubsection{A class of models}

A large class of models has been constructed in \cite{ssw, ssw3}  by considering a Lie algebra $\mathfrak{g}$ and a representation ${\cal R}$. The bosonic fields of the
 vector and tensor multiplets are chosen as
\bea
A^r=(A^\km, A^A)~,~~~Y^r=(Y^\km, Y^A)~,~~~B^I=(B^A, B_A)~,~~~\phi^I=(\phi^A, \phi_A)~,
\eea
ie $A$ and $Y$ take values in $\mathfrak{g}\oplus {\cal R}$ while $B$ and $\phi$ take values in  ${\cal R}\oplus {\cal R}^* $.  Moreover the
non-vanishing couplings are chosen as
\bea
&&\eta^A{}_B=\eta_B{}^A=\delta^A_B~,~~~h^B{}_A=g_A{}^B=\delta^B_A~,~~~f_{\km A}{}^B=-{1\over2} (T_\km)_A{}^B~,~~~f_{\km\kn}{}^\kp~,
\cr
&&d^B{}_{\km A}={1\over2} b^B{}_{A\km}={1\over2} b^B{}_{\km A}={1\over2} (T_\km)_A{}^B~,~~~d_{ABC}=d_{(ABC)}=b_{BCA}~,~~~
\cr
&&d_{AB\km}=d_{(AB)\km}={1\over2} b_{AB\km}={1\over2} b_{A\km B}~,~~d_{A\km\kn}~,~~
b_{A(\km\kn)}=2 d_{A(\km\kn)}~,~~\theta_\km{}^\kn=\delta_\km{}^\kn~,
\eea
where $T_\km$ are the representation matrices of $\mathfrak{g}$ in ${\cal R}$. These solve all the constraints on the couplings imposed on these models
provided that $d_{\km AB}, d_{\km\kn A}$ and $d_{ABC}$ are invariant under the action of $\mathfrak{g}$.

\subsubsection{Self-dual string solitons from instantons}

Motivated from the M-brane intersection rules, we shall seek   self-dual string solitons in the  class of models described in the previous section which preserve $1/2$ of the supersymmetry. The relevant class of
supersymmetric backgrounds for self-dual string solitons are those with 4 Killing spinors that have isotropy group $Sp(1)\ltimes \bH$ in table 4.  The conditions
on the fields of the vector and tensor multiplets are given in \cite{ap3} and in section 3.6 for the hyper-multiplet scalars. Similar solutions have been found in \cite{ap3} for another class of
models, see also \cite{palmer}. The self-dual string soliton on a single M5-brane has been found in \cite{howe} and it is singular at the position of the string.

To solve the supersymmetry conditions, Bianchi identities and field equations, suppose that the fields have support on 4-directions transverse to the light-cone coordinates $(x^+, x^-)$ which are identified with the world-sheet of the string. In addition  choose
\bea
\mathcal{F}^r=(\mathcal{F}^\km,0)~,~~\mathcal{H}^I=(0, \mathcal{H}_A)~,~~\phi^I=(0, \phi_A)~,~~\mathcal{H}_r^{(4)}=Y^r=\mathcal{H}^{(5)}=0~,
\eea
with $\mathcal{F}^r$ purely magnetic.
We focus on models for which the only non-vanishing coupling constants with all indices lowered are $b_{A\km\kn}, d_{A\km\kn}$.  In addition we assume that either the model is not coupled
to hyper-multiplets or if it is coupled, then the hyper-scalars are at a maximally supersymmetric vacuum for consistency, ie the gauging and the hyper-K\"ahler cone has been chosen such that there is a value $q=q_0$ and
\bea
\mu_\km(q_0)=0~,~~~\partial_\tM\mu_\km(q_0)=0~,
\eea
 where $\mu$ are the moment maps defined in (\ref{moment}).
For the flat hyperk\"ahler cone, such a value is $q_0=0$ or any other fixed point of rotational isometries that are gauged. In either case, the contribution
from the hyper-multiplets decouples.

The remaining  non-trivial Bianchi identities and field equations that one has to demonstrate are
\bea
D_{[m}\mathcal{F}^\km_{n\ell]}=0~,~~~b^B{}_{A\kn }\mathcal{F}^\kn_{m\ell} \phi_B=0~,~~~
d^B{}_{A\mathfrak{n}} \mathcal{F}^\kn_{[\mu\nu} \mathcal{H}_{\lambda\rho\sigma ]B}=0~,~~~\la{cona}
\eea
and
\bea
D_{[\mu} \mathcal {H}_{\nu\rho\sigma] A}={3\over2} d_{A\km\kn} \mathcal{F}^\km_{[\mu\nu} \mathcal{F}^\kn_{\rho\sigma]}~,  D_m D^m\phi_A=-{1\over2} d_{A\km\kn}  \mathcal{F}^\km_{mn}  \mathcal{F}^{\kn mn}~.
\la{conb}
\eea

These conditions can be solved provided that ${\cal R}$ can be decomposed as ${\cal R}=I\oplus {\cal R}'$, where $I$ is a trivial representation of $\mathfrak{g}$ and take that
$\phi_A$ and $\mathcal{H}_A$ vanish unless they lie along the trivial representation, and denote the non-vanishing fields with $\phi_0$ and $\mathcal{H}_0$, respectively. Such a choice will solve the last two conditions in (\ref{cona}) as $T_\km$ vanishes
along the trivial representation. The first condition in (\ref{cona}) is solved by identifying $\mathcal{F}^\km$ with the field strength of a gauge field with Lie algebra $\mathfrak{g}$.

It remains to solve the conditions in (\ref{conb}). First observe that $D_m D^m\phi_0=\partial_m\partial^m \phi_0$, and similarly on $\mathcal{H}_0$, and identify $d_{0\km\kn}$ with a bi-invariant metric on $\mathfrak{g}$.   Next set
\bea
\mathcal{H}_0=- \partial_i\phi_0\, dx^-\wedge dx^+\wedge dx^i+{1\over3!}
\partial_j\phi_0\,\epsilon^j{}_{i_1 i_2 i_3}\,\,dx^{i_1}\wedge dx^{i_2}\wedge dx^{i_3} ~,~~
\la{exh}
\eea
Then recall that the KSEs for $Sp(1)\ltimes \bH$ invariant spinors
require that $\mathcal{F}$ is an anti-self dual instanton. Because of this and (\ref{exh}), the second condition in (\ref{conb}) implies the first.  Finally, the last condition
in (\ref{conb}) is solved because the Pontryagin form of instantons can be written as the Laplacian on a scalar function \cite{osborn}.  In addition for generic values of instanton moduli space, all the string solutions
are smooth.

To present an explicit solution take $\mathfrak{g}=\mathfrak{su}(2)$, we consider the  configuration  with instanton number 1 and use the results of \cite{halfhet}. In such a case, the gauge connection $A$ of $\mathcal{F}^\km=\mathcal{F}^{ab}$ and $\phi_0$
can be written as
\bea
A^{ab}&=&2 (J^{r'})^{ab} (J_{r'})_{ij} {x^j\over |x|^2+\rho^2}\,\, e^i~,~~~\phi_0=c+ 4\sqrt 2{|x|^2+2\rho^2\over (|x|^2+\rho^2)^2}+ h_0~,~~~
\cr
h_0&=&\sum_\nu{Q_\nu\over |x-x_\nu|^2}~,
\eea
where $x$ are the coordinates in $\bR^{5,1}$ transverse to string worldsheet coordinates  $(x^+, x^-)$, $c$ is a constant, and $\rho$ is the instanton modulus. Moreover   $h_0$ is a multi-centred harmonic function, which if it is included in the solution, then delta function sources have to be added in the field equation for $\phi_0$.
Let us focus on the solution with $h_0=0$.  Such a solution is smooth at a generic value of $\rho$. At large $|x|$, ie far away from the string, the scalar $\phi_0$ converges to the constant $c$, and the gauge connection is a pure gauge. As $|x|$ becomes small,
the values of $\phi_0$ and $A$ are regulated by the modulus $\rho\not=0$ of the instanton. In particular at $|x|=0$, the value
of $\phi_0$ is $c+(8\sqrt 2/\rho^2)$. Assuming that the theory describes a M5-brane, $c$ becomes the position of M5 at infinity. Then the M5-brane is ``pulled'' by the M2-branes ending on it and its position shifts by  $8\sqrt 2/\rho^2$. Of course as the instanton size
becomes small, $\rho^2\rightarrow 0$,  a throat is developed. This solution becomes similar to self-dual strings of \cite{howe}.

The dyonic string charge $q_s$ of all solutions can be computed by integrating $\mathcal{H}_0$ on the 3-sphere at infinity. After an appropriate normalization, this can be
identified with the instanton number $k$, ie
\bea
q_s=\int_{S^3\subset \bR^4} \mathcal{H}_0=k~.
\eea
All  solutions with any instanton number $k$ are smooth at a generic point in the instanton moduli space.

\subsection{3-branes}

Motivated from the M-brane intersection rules which state that two M5-branes intersect on a 3-brane, we shall describe a class of models which exhibit 3-brane solitons. These are those for which all the potentials vanish and the only active fields are those of the hyper-multiplets.
Moreover, the hyper-multiplet scalars depend only on the two transverse directions to the 3-brane soliton. First to identify the models with 3-brane solitons
suppose that the hyper-multiplets are not gauged, ie the embedding tensor $\theta=0$.  Moreover set all the fields apart from the hypermultiplet scalars $q$
and $\mathcal{H}^{(5)}$ equal to zero. The only non-trivial conditions that have to be satisfied to construct solutions are the field equations for $q$
and the hyperini KSEs.

To solve the hypernini KSEs, we  take the case with 4 supersymmetries and compact isotropy group.  The relevant equations are given in  (\ref{n4compb}).  The solution of KSEs implies that the hyper-multiplet scalars do not depend on four directions, as expected for a 3-brane soliton, and (\ref{n4compb})  is a   Cauchy-Riemann
equations which implies  that $q$ is a holomorphic curve into the hyper-K\"ahler cone.  In addition, the field equation for the $q$'s  is automatically satisfied.

Utilizing the $N=2$ solutions with compact isotropy group, a similar argument reveals   the existence of string solitons preserving 1/4 of supersymmetry supported by a holomorphic surface embedded
into the hyper-K\"ahler cone. It is expected that such solitons are  associated with a triple M5-brane intersection on a string.

\newsection{Conclusions}

A distinct  role  amongst the solutions of a supersymmetric theory have those that preserve some of the supercharges. Such solutions apart from the field equations also solve
the KSEs of supersymmetric theories. In the context of string theory, M-theory and supergravity such solutions
have found widespread applications to compactifications, black holes, AdS/CFT, and branes.
They have also been instrumental in understanding string dualities. The systematic investigation of supersymmetric solutions
is an outstanding problem and  is instrumental in the development of various aspects of string and M-theory
as these require a deeper understanding of such solutions.  Apart from the applications to physics, there are notable
applications to geometry as intricate geometric structures arise in the description of such solutions.

Spinorial geometry provides a general framework to understand the solution of the KSEs of supersymmetric systems.
It has been used to systematically solve the KSEs of heterotic supergravity, the KSEs of $D=4$ ${\cal N}=1$ supergravity and those of (1,0) 6-dimensional
supergravity  to determine both the fractions of supersymmetry preserved and the geometries of all backgrounds.
It can also be used to solve the KSEs of  supersymmetric theories for a small or near maximal number of supersymmetries.

In this review, the spinorial geometry method has been described as it applies in the 6-dimensional (1,0) supergravity.
It has been explained how all the fractions of supersymmetry preserved by the supersymmetric
backgrounds have been identified as well as how the KSEs can be solved to determine the conditions
on the fields and the spacetime geometry. In addition two applications have been presented. One is
on the near horizon geometries of 6-dimensional black holes.  In particular, it is explained how a class of such
horizons is locally a product $AdS_3\times \Sigma^3$. Another application is on the description of the brane
 solitons of 6-dimensional (1,0) superconformal theories. In particular a systematic description of
 all configurations that preserve a fraction of supersymmetry is given.

The applicability of spinorial geometry is not limited to six dimensions. It can be applied to supersymmetric systems in all dimensions
providing a systematic way to identify the supersymmetric backgrounds. It is expected that in the next few years
a clear picture will emerge of the geometry of all such solutions. Applications will include insights into the
backgrounds used in AdS/CFT, the discovery of new black holes in various dimensions and the unraveling of their symmetries,
the understanding of brane solutions and their intersections, and the exploration of superconformal  theories.

\vskip 1.0cm
\noindent{\bf Acknowledgements} \vskip 0.1cm
\noindent
I would like to thank the Albert-Einstein-Institute,  Max Planck Institute  in  Golm,  for providing  a stimulating environment
to complete this review. I am partially supported by the STFC grant ST/J002798/1.


\vskip 0.5cm







\begin{thebibliography}{99}


\bibitem{spingeom}
  J.~Gillard, U.~Gran and G.~Papadopoulos,
  ``The spinorial geometry of supersymmetric backgrounds,''
  Class.\ Quant.\ Grav.\  {\bf 22} (2005) 1033
  [arXiv:hep-th/0410155].

  \bibitem{manton}
  N.~Manton and P.~Sutcliffe, ``Topological Solitons,'' CUP (2004), Cambridge.

 \bibitem{olive}
  C.~Montonen and D.~I.~Olive,
  ``Magnetic Monopoles as Gauge Particles?,''
  Phys.\ Lett.\ B {\bf 72} (1977) 117.





\bibitem{witten}
  N.~Seiberg and E.~Witten,
  ``Electric - magnetic duality, monopole condensation, and confinement in N=2 supersymmetric Yang-Mills theory,''
  Nucl.\ Phys.\ B {\bf 426} (1994) 19
   [Erratum-ibid.\ B {\bf 430} (1994) 485]
  [hep-th/9407087].


  \bibitem{duff}
  M.~J.~Duff, B.~E.~W.~Nilsson and C.~N.~Pope,
  ``Kaluza-Klein Supergravity,''
  Phys.\ Rept.\  {\bf 130} (1986) 1.


  \bibitem{maeda}
  K.~-i.~Maeda and M.~Nozawa,
  ``Black hole solutions in string theory,''
  Prog.\ Theor.\ Phys.\ Suppl.\  {\bf 189} (2011) 310
  [arXiv:1104.1849 [hep-th]].



\bibitem{stelle}
  K.~S.~Stelle,
  ``Lectures on supergravity p-branes,''
  In *Trieste 1996, High energy physics and cosmology* 287-339
  [hep-th/9701088].


  \bibitem{smith}
  D.~J.~Smith,
  ``Intersecting brane solutions in string and M theory,''
  Class.\ Quant.\ Grav.\  {\bf 20} (2003) R233
  [hep-th/0210157].


  \bibitem{obers}
  N.~A.~Obers and B.~Pioline,
  ``U duality and M theory,''
  Phys.\ Rept.\  {\bf 318} (1999) 113
  [hep-th/9809039].




  \bibitem{maldacena}
  O.~Aharony, S.~S.~Gubser, J.~M.~Maldacena, H.~Ooguri and Y.~Oz,
  ``Large N field theories, string theory and gravity,''
  Phys.\ Rept.\  {\bf 323} (2000) 183
  [hep-th/9905111].


  \bibitem{tod}
  K.~P.~Tod,
  ``All Metrics Admitting Supercovariantly Constant Spinors,''
  Phys.\ Lett.\ B {\bf 121} (1983) 241.


  \bibitem{hull}
  J.~P.~Gauntlett, J.~B.~Gutowski, C.~M.~Hull, S.~Pakis and H.~S.~Reall,
  ``All supersymmetric solutions of minimal supergravity in five- dimensions,''
  Class.\ Quant.\ Grav.\  {\bf 20} (2003) 4587
  [hep-th/0209114].


  \bibitem{pakis}
  J.~P.~Gauntlett and S.~Pakis,
  ``The Geometry of D = 11 killing spinors,''
  JHEP {\bf 0304} (2003) 039
  [hep-th/0212008].

   J.~P.~Gauntlett, J.~B.~Gutowski and S.~Pakis,
  ``The Geometry of D = 11 null Killing spinors,''
  JHEP {\bf 0312} (2003) 049
  [hep-th/0311112].

 \bibitem{gpjose}
  J.~M.~Figueroa-O'Farrill and G.~Papadopoulos,
  ``Maximally supersymmetric solutions of ten-dimensional and eleven-dimensional supergravities,''
  JHEP {\bf 0303} (2003) 048
  [hep-th/0211089].

  \bibitem{systway}
  U.~Gran, G.~Papadopoulos and D.~Roest,
  ``Systematics of M-theory spinorial geometry,''
  Class.\ Quant.\ Grav.\  {\bf 22} (2005) 2701
  [hep-th/0503046].


  U.~Gran, J.~Gutowski, G.~Papadopoulos and D.~Roest,
  ``Systematics of IIB spinorial geometry,''
  Class.\ Quant.\ Grav.\  {\bf 23} (2006) 1617
  [hep-th/0507087].

  \bibitem{iibgrangp}
  U.~Gran, J.~Gutowski and G.~Papadopoulos,
  ``The Spinorial geometry of supersymmetric IIb backgrounds,''
  Class.\ Quant.\ Grav.\  {\bf 22} (2005) 2453
  [hep-th/0501177].

  U.~Gran, J.~Gutowski and G.~Papadopoulos,
  ``The G(2) spinorial geometry of supersymmetric IIB backgrounds,''
  Class.\ Quant.\ Grav.\  {\bf 23} (2006) 143
  [hep-th/0505074].

 \bibitem{iiagrangp}
  U.~Gran, G.~Papadopoulos and C.~von Schultz,
  ``Supersymmetric geometries of IIA supergravity I,''
  arXiv:1401.6900 [hep-th].


  \bibitem{het1}
  U.~Gran, P.~Lohrmann and G.~Papadopoulos,
  ``The spinorial geometry of supersymmetric heterotic string backgrounds,''
  JHEP {\bf 0602} (2006) 063
  [arXiv:hep-th/0510176].

\bibitem{het2}
  U.~Gran, G.~Papadopoulos, D.~Roest and P.~Sloane,
  ``Geometry of all supersymmetric type I backgrounds,''
  JHEP {\bf 0708} (2007) 074
  [arXiv:hep-th/0703143].

 U.~Gran, G.~Papadopoulos and D.~Roest,
  ``Supersymmetric heterotic string backgrounds,''
  Phys.\ Lett.\  B {\bf 656} (2007) 119
  [arXiv:0706.4407 [hep-th]].


\bibitem{ap1}
  M.~Akyol and G.~Papadopoulos,
  ``Spinorial geometry and Killing spinor equations of 6-D supergravity,''
  Class.\ Quant.\ Grav.\  {\bf 28} (2011) 105001
  [arXiv:1010.2632 [hep-th]].

  \bibitem{grover1}
  J.~Grover, J.~B.~Gutowski and W.~Sabra,
  ``Null Half-Supersymmetric Solutions in Five-Dimensional Supergravity,''
  JHEP {\bf 0810} (2008) 103
  [arXiv:0802.0231 [hep-th]].

  \bibitem{d4grangp}
  U.~Gran, J.~Gutowski and G.~Papadopoulos,
  ``Geometry of all supersymmetric four-dimensional N = 1 supergravity backgrounds,''
  JHEP {\bf 0806} (2008) 102
  [arXiv:0802.1779 [hep-th]].

  \bibitem{klemm1}
  S.~L.~Cacciatori, D.~Klemm, D.~S.~Mansi and E.~Zorzan,
  ``All timelike supersymmetric solutions of N=2, D=4 gauged supergravity coupled to abelian vector multiplets,''
  JHEP {\bf 0805} (2008) 097
  [arXiv:0804.0009 [hep-th]].

\bibitem{grover2}
  J.~Grover, J.~B.~Gutowski, C.~A.~R.~Herdeiro and W.~Sabra,
  ``HKT Geometry and de Sitter Supergravity,''
  Nucl.\ Phys.\ B {\bf 809} (2009) 406
  [arXiv:0806.2626 [hep-th]].

\bibitem{klemm2}
D.~Klemm and E.~Zorzan,
``All null supersymmetric backgrounds of N=2, D=4 gauged supergravity coupled to abelian vector multiplets,''
  Class.\ Quant.\ Grav.\  {\bf 26} (2009) 145018
[arXiv:0902.4186 [hep-th]].


\bibitem{iibnearmax}
  U.~Gran, J.~Gutowski, G.~Papadopoulos and D.~Roest,
  ``N=31 is not IIB,''
  JHEP {\bf 0702} (2007) 044
  [hep-th/0606049].

   U.~Gran, J.~Gutowski, G.~Papadopoulos and D.~Roest,
  ``IIB solutions with N > 28 Killing spinors are maximally supersymmetric,''
  JHEP {\bf 0712} (2007) 070
  [arXiv:0710.1829 [hep-th]].

  U.~Gran, J.~Gutowski and G.~Papadopoulos,
  ``Classification of IIB backgrounds with 28 supersymmetries,''
  JHEP {\bf 1001} (2010) 044
  [arXiv:0902.3642 [hep-th]].


  \bibitem{d11nearmax}
  U.~Gran, J.~Gutowski, G.~Papadopoulos and D.~Roest,
  ``N=31, D=11,''
  JHEP {\bf 0702} (2007) 043
  [hep-th/0610331].


   U.~Gran, J.~Gutowski and G.~Papadopoulos,
  ``M-theory backgrounds with 30 Killing spinors are maximally supersymmetric,''
  JHEP {\bf 1003} (2010) 112
  [arXiv:1001.1103 [hep-th]].

  \bibitem{sezgin}
  H.~Nishino and E.~Sezgin,
  ``Matter And Gauge Couplings Of N=2 Supergravity In Six-Dimensions,''
  Phys.\ Lett.\  B {\bf 144} (1984) 187.

  ``The Complete N=2, D = 6 Supergravity With Matter And Yang-Mills
  Couplings,''
  Nucl.\ Phys.\  B {\bf 278} (1986) 353.

  ``New couplings of six-dimensional supergravity,''
  Nucl.\ Phys.\  B {\bf 505} (1997) 497
  [arXiv:hep-th/9703075].

\bibitem{ferrara}
  S.~Ferrara, F.~Riccioni and A.~Sagnotti,
  ``Tensor and vector multiplets in six-dimensional supergravity,''
  Nucl.\ Phys.\  B {\bf 519} (1998) 115
  [arXiv:hep-th/9711059].

\bibitem{salamon}
S.~Salamon, ``Quaternionic K\"ahler Manifolds,'' Invent. Math. {\bf 67} (1982), 143.


\bibitem{riccioni}
  F.~Riccioni,
  ``All couplings of minimal six-dimensional supergravity,''
  Nucl.\ Phys.\  B {\bf 605} (2001) 245
  [arXiv:hep-th/0101074].



\bibitem{ap2}
  M.~Akyol and G.~Papadopoulos,
  ``Topology and geometry of 6-dimensional (1,0) supergravity black hole horizons,''
  Class.\ Quant.\ Grav.\  {\bf 29} (2012) 055002
  [arXiv:1109.4254 [hep-th]].


  \bibitem{ap3}
  M.~Akyol and G.~Papadopoulos,
  ``(1,0) superconformal theories in six dimensions and Killing spinor equations,''
  JHEP {\bf 1207} (2012) 070
  [arXiv:1204.2167 [hep-th]].

\bibitem{ap4}
 M.~Akyol and G.~Papadopoulos,
  ``Brane solitons of (1,0) superconformal theories in six dimensions with hypermultiplets,''
  arXiv:1307.1041 [hep-th].

\bibitem{ssw}
  H.~Samtleben, E.~Sezgin and R.~Wimmer,
  ``(1,0) superconformal models in six dimensions,''
  JHEP {\bf 1112} (2011) 062
  [arXiv:1108.4060 [hep-th]].

  H.~Samtleben, E.~Sezgin, R.~Wimmer and L.~Wulff,
  ``New superconformal models in six dimensions: Gauge group and representation structure,''
  PoS CORFU2011 (2011) 071
  [arXiv:1204.0542 [hep-th]].

\bibitem{ssw3}
	H.~Samtleben, E.~Sezgin and R.~Wimmer,
  ``Six-dimensional superconformal couplings of non-abelian tensor and hypermultiplets,''
   [arXiv:1212.5199 [hep-th]].


 \bibitem{strominger}
  A.~Strominger,
  ``Open p-branes,''
  Phys.\ Lett.\ B {\bf 383} (1996) 44
  [hep-th/9512059].

  \bibitem{pktgp}
  G.~Papadopoulos and P.~K.~Townsend,
  ``Intersecting M-branes,''
  Phys.\ Lett.\ B {\bf 380} (1996) 273
  [hep-th/9603087].

\bibitem{gal}
K.~ Galicki, ``A generalization of the momentum mapping
construction for quaternionic Kähler manifolds,'' Commun.\
Math.\ Phys.\  {\bf 108} (1987) 117.


\bibitem{dario}
  J.~B.~Gutowski, D.~Martelli and H.~S.~Reall,
  ``All supersymmetric solutions of minimal supergravity in six dimensions,''
  Class.\ Quant.\ Grav.\  {\bf 20} (2003) 5049
  [arXiv:hep-th/0306235].


\bibitem{jose}
  A.~Chamseddine, J.~M.~Figueroa-O'Farrill and W.~Sabra,
  ``Supergravity vacua and Lorentzian Lie groups,''
  arXiv:hep-th/0306278.

  \bibitem{han}
  M.~Cariglia and O.~A.~P.~Mac Conamhna,
  ``The general form of supersymmetric solutions of N = (1,0) U(1) and  SU(2)
  gauged supergravities in six dimensions,''
  Class.\ Quant.\ Grav.\  {\bf 21} (2004) 3171
  [arXiv:hep-th/0402055].

\bibitem{jong}
  D.~C.~Jong, A.~Kaya and E.~Sezgin,
  ``6D dyonic string with active hyperscalars,''
  JHEP {\bf 0611} (2006) 047
  [arXiv:hep-th/0608034].

\bibitem{gueven}
  R.~Gueven, J.~T.~Liu, C.~N.~Pope and E.~Sezgin,
  ``Fine tuning and six-dimensional gauged N = (1,0) supergravity vacua,''
  Class.\ Quant.\ Grav.\  {\bf 21} (2004) 1001
  [arXiv:hep-th/0306201].

 \bibitem{qkt}
  P.~S.~Howe, A.~Opfermann and G.~Papadopoulos,
  ``Twistor spaces for QKT manifolds,''
  Commun.\ Math.\ Phys.\  {\bf 197} (1998) 713
  [arXiv:hep-th/9710072].



\bibitem{medina}
  A.~Medina and P.~Revoy, ``Algebres de Lie et produit scalaire invariant'', Ann. Scient. Ec. Norm. Sup. {\bf 18}
  (1985) 553.

  \bibitem{josec}
   T.~Kawano and S.~Yamaguchi,
  ``Dilatonic parallelizable NS-NS backgrounds,''
  Phys.\ Lett.\  B {\bf 568} (2003) 78
  [arXiv:hep-th/0306038].

   J.~M.~Figueroa-O'Farrill, T.~Kawano, S.~Yamaguchi,
  ``Parallelizable heterotic backgrounds,''
  JHEP {\bf 0310 } (2003)  012.
  [hep-th/0308141].


\bibitem{israel}
  W.~Israel,
  ``Event Horizons In Static Vacuum Space-Times,''
  Phys.\ Rev.\  {\bf 164} (1967) 1776.

\bibitem{carter}
  B.~Carter,
  ``Axisymmetric Black Hole Has Only Two Degrees of Freedom,''
  Phys.\ Rev.\ Lett.\  {\bf 26} (1971) 331.


\bibitem{hawking}
 S.~W.~Hawking,
  ``Black holes in general relativity,''
  Commun.\ Math.\ Phys.\  {\bf 25} (1972) 152.


\bibitem{robinson1}
  D.~C.~Robinson,
  ``Uniqueness of the Kerr black hole,''
  Phys.\ Rev.\ Lett.\  {\bf 34} (1975) 905.

\bibitem{israel2}
W. Israel, ``Event Horizons in Static, Electrovac Space-Times,"
Commun. Math. Phys. {\bf{8}} (1968) 245.

\bibitem{mazur}
P. O. Mazur, ``Proof of Uniqueness of the Kerr-Newman Black Hole Solution,"
J. Phys. A {\bf{15}} (1982) 3173.





\bibitem{robinson} D.~ Robinson, ``Four decades of black hole uniqueness theorems,'' appeared
in {\it The Kerr spacetime: Rotating black holes in General Relativity}, eds D.~L.~ Wiltshire, M.~ Visser and
S.~ M.~ Scott, pp 115-143,  CUP 2009.

\bibitem{bmpv}
J. C. Breckenridge, R. C. Myers, A. W. Peet and C. Vafa,
``D-branes and spinning black holes",
Phys. Lett. {\bf{B391}} (1997) 93; hep-th/9602065.

\bibitem{reallbh}
H. S. Reall,
``Higher dimensional black holes and supersymmetry",
Phys. Rev. {\bf{D68}} (2003) 024024; hep-th/0211290.




\bibitem{ring1}
H. Elvang, R. Emparan, D. Mateos and H. S. Reall,
``A Supersymmetric black ring", Phys. Rev. Lett. {\bf{93}} (2004) 211302;  hep-th/0407065.



\bibitem{gibbons1}
G. W. Gibbons, D. Ida and T. Shiromizu,
``Uniqueness and non-uniqueness of
static black holes in higher dimensions",
Phys. Rev. Lett. {\bf{89}} (2002) 041101;
hep-th/0206049.

\bibitem{rogatko}
M. Rogatko,
``Uniqueness theorem of static degenerate and non-degenerate
charged black holes in higher dimensions", Phys. Rev. {\bf{D67}} (2003) 084025;
hep-th/0302091; ``Classification of static charged black holes in higher dimensions,''
Phys. Rev. {\bf{D73}} (2006), 124027; hep-th/0606116.

\bibitem{reall3}
  H.~K.~Kunduri, J.~Lucietti and H.~S.~Reall,
  ``Near-horizon symmetries of extremal black holes,''
  Class.\ Quant.\ Grav.\  {\bf 24 } (2007)  4169; arXiv:0705.4214 [hep-th].

\bibitem{obers1}
R. Emparan, T. Harmark, V. Niarchos and N. Obers,
``World-Volume Effective Theory for Higher-Dimensional Black Holes,"
Phys. Rev. Lett. {\bf{102}} (2009) 191301;
 arXiv:0902.0427 [hep-th];
 ``Essentials of Blackfold Dynamics;" arXiv:0910.1601 [hep-th].


\bibitem{kunduri}
  H.~K.~Kunduri and J.~Lucietti,
  ``An infinite class of extremal horizons in higher dimensions,''
  Commun.\ Math.\ Phys.\  {\bf 303} (2011) 31
  [arXiv:1002.4656 [hep-th]]; 
  ``Extremal Sasakian horizons,''
  Phys.\ Lett.\ B {\bf 713} (2012) 308
  [arXiv:1204.5149 [hep-th]];
   ``Degenerate horizons, Einstein metrics, and Lens space bundles,''
  arXiv:1210.1268 [hep-th].
  
 \bibitem{bhreview}
  H.~K.~Kunduri and J.~Lucietti,
  ``Classification of near-horizon geometries of extremal black holes,''
  Living Rev.\ Rel.\  {\bf 16} (2013) 8
  [arXiv:1306.2517 [hep-th]].

\bibitem{gt}
  G.~W.~Gibbons and P.~K.~Townsend,
  ``Vacuum interpolation in supergravity via super p-branes,''
  Phys.\ Rev.\ Lett.\  {\bf 71} (1993) 3754
  [hep-th/9307049].


\bibitem{jgdm} J.~B.~Gutowski, D.~Martelli and H.~S.~Reall,
  ``All supersymmetric solutions of minimal supergravity in six dimensions,''
  Class.\ Quant.\ Grav.\  {\bf 20} (2003) 5049
  [arXiv:hep-th/0306235].


  \bibitem{fourhor} J.~Gutowski and G.~Papadopoulos,
  ``Topology of supersymmetric N=1, D=4 supergravity horizons,''
  JHEP {\bf 1011} (2010) 114
  [arXiv:1006.4369 [hep-th]].


 \bibitem{hh} J.~Gutowski, G.~Papadopoulos,
  ``Heterotic Black Horizons,''
  JHEP {\bf 07}, 011 (2010); arXiv:0912.3472 [hep-th].

J.~Gutowski and G.~Papadopoulos,
  ``Heterotic horizons, Monge-Ampere equation and del Pezzo surfaces,''
  JHEP {\bf 1010} (2010) 084
  [arXiv:1003.2864 [hep-th]].


\bibitem{iibhor}  U.~Gran, J.~Gutowski, G.~Papadopoulos,
  ``IIB black hole horizons with five-form flux and KT geometry,''
  JHEP {\bf 05 } (2011)  050; arXiv:1101.1247 [hep-th].

 U.~Gran, J.~Gutowski and G.~Papadopoulos,
  ``IIB black hole horizons with five-form flux and extended supersymmetry,''
  JHEP {\bf 1109} (2011) 047
  [arXiv:1104.2908 [hep-th]].

  U.~Gran, J.~Gutowski and G.~Papadopoulos,
  ``IIB horizons,''
  Class.\ Quant.\ Grav.\  {\bf 30} (2013) 205004
  [arXiv:1304.6539 [hep-th]].



\bibitem{mhor}  J.~Gutowski and G.~Papadopoulos,
  ``Static M-horizons,''
  JHEP {\bf 1201} (2012) 005
  [arXiv:1106.3085 [hep-th]].

   J.~Gutowski and G.~Papadopoulos,
  ``M-Horizons,''
  JHEP {\bf 1212} (2012) 100
  [arXiv:1207.7086].


 \bibitem{index}
  J.~Grover, J.~B.~Gutowski, G.~Papadopoulos and W.~A.~Sabra,
  ``Index Theory and Supersymmetry of 5D Horizons,''
  arXiv:1303.0853 [hep-th].


J.~Gutowski and G.~Papadopoulos,
  ``Index theory and dynamical symmetry enhancement of M-horizons,''
  JHEP {\bf 1305} (2013) 088
  [arXiv:1303.0869 [hep-th]].


 U.~Gran, J.~Gutowski and G.~Papadopoulos,
  ``Index theory and dynamical symmetry enhancement near IIB horizons,''
  JHEP {\bf 1311} (2013) 104
  [arXiv:1306.5765 [hep-th]].



\bibitem{groverrings}
  J.~Grover, J.~B.~Gutowski and W.~A.~Sabra,
  ``Supersymmetric AdS Black Rings,''
  arXiv:1306.0017 [hep-th].




\bibitem{ringkun}
  H.~K.~Kunduri, J.~Lucietti and H.~S.~Reall,
  ``Do supersymmetric anti-de Sitter black rings exist?,''
  JHEP {\bf 0702} (2007) 026
  [hep-th/0611351].
  
  \bibitem{wald}
  V.~Moncrief and J.~Isenberg,
  ``Symmetries of cosmological Cauchy horizons,''
  Commun.\ Math.\ Phys.\  {\bf 89} (1983) 3,  387.


H. Friedrich, I. Racz and R. M. Wald,
``On the rigidity theorem for space-times with a stationary event horizon or a compact Cauchy horizon,''
Commun. Math. Phys. {\bf{204}} (1999) 691; gr-qc/9811021.



\bibitem{hkt} P.~S.~Howe and G.~Papadopoulos,
  ``Twistor spaces for HKT manifolds,''
  Phys.\ Lett.\  B {\bf 379} (1996) 80
  [arXiv:hep-th/9602108].


\bibitem{ppetersen} P.~ Petersen, ``Riemannian Geometry,'' Graduate Texts in Mathematics, Springer (1998), page 237.

\bibitem{poincare} G.~ Perelman, ``The entropy formula for the Ricci flow and its geometric applications,''
[arXiv:math/0211159]; ``Ricci flow with surgery on three-manifolds,''
    [ arXiv:math/0303109]; ``Finite extinction time for the solutions to the Ricci flow on certain three-manifolds,''  [arXiv:math/0307245]



\bibitem{ferrara2}
L.~Andrianopoli, S.~Ferrara, A.~Marrani and M.~Trigiante,
  ``Non-BPS Attractors in 5d and 6d Extended Supergravity,''
  Nucl.\ Phys.\ B {\bf 795} (2008) 428
  [arXiv:0709.3488 [hep-th]].

  S.~Ferrara, A.~Marrani, J.~F.~Morales and H.~Samtleben,
  ``Intersecting Attractors,''
  Phys.\ Rev.\ D {\bf 79} (2009) 065031
  [arXiv:0812.0050 [hep-th]].





\bibitem{lp}
  N.~Lambert and C.~Papageorgakis,
  ``Nonabelian (2,0) Tensor Multiplets and 3-algebras,''
  JHEP {\bf 1008} (2010) 083
  [arXiv:1007.2982 [hep-th]].

\bibitem{chu}
  C.~-S.~Chu,
  ``A Theory of Non-Abelian Tensor Gauge Field with Non-Abelian Gauge Symmetry G x G,''
  arXiv:1108.5131 [hep-th].

 \bibitem{chu2}
  C.~-S.~Chu and S.~-L.~Ko,
  ``Non-abelian Action for Multiple M5-Branes,''
  arXiv:1203.4224 [hep-th].

  \bibitem{gp}
   J.~M.~Figueroa-O'Farrill and G.~Papadopoulos,
  ``Plucker type relations for orthogonal planes,''
  math/0211170 [math-ag].


  G.~Papadopoulos,
  ``M2-branes, 3-Lie Algebras and Plucker relations,''
  JHEP {\bf 0805} (2008) 054
  [arXiv:0804.2662 [hep-th]].

  \bibitem{gg}
  J.~P.~Gauntlett and J.~B.~Gutowski,
  ``Constraining Maximally Supersymmetric Membrane Actions,''
  JHEP {\bf 0806} (2008) 053
  [arXiv:0804.3078 [hep-th]].


\bibitem{douglas}
  M.~R.~Douglas,
  ``On D=5 super Yang-Mills theory and (2,0) theory,''
  JHEP {\bf 1102} (2011) 011
  [arXiv:1012.2880 [hep-th]].  See also the talk of the author at Strings 2013.

\bibitem{m5h}
  P.~S.~Howe and E.~Sezgin,
  ``D = 11, p = 5,''
  Phys.\ Lett.\ B {\bf 394} (1997) 62
  [hep-th/9611008].


\bibitem{m5s}
  P.~Pasti, D.~P.~Sorokin and M.~Tonin,
  ``Covariant action for a D = 11 five-brane with the chiral field,''
  Phys.\ Lett.\ B {\bf 398} (1997) 41
  [hep-th/9701037].

  \bibitem{m5sh}
  M.~Aganagic, J.~Park, C.~Popescu and J.~H.~Schwarz,
  ``World volume action of the M theory five-brane,''
  Nucl.\ Phys.\ B {\bf 496} (1997) 191
  [hep-th/9701166].




 \bibitem{bl}
  J.~Bagger and N.~Lambert,
  ``Gauge symmetry and supersymmetry of multiple M2-branes,''
  Phys.\ Rev.\ D {\bf 77} (2008) 065008
  [arXiv:0711.0955 [hep-th]].

  \bibitem{gust}
  A.~Gustavsson,
  ``Algebraic structures on parallel M2-branes,''
  Nucl.\ Phys.\ B {\bf 811} (2009) 66
  [arXiv:0709.1260 [hep-th]].

\bibitem{abjm}
  O.~Aharony, O.~Bergman, D.~L.~Jafferis and J.~Maldacena,
  ``N=6 superconformal Chern-Simons-matter theories, M2-branes and their gravity duals,''
  JHEP {\bf 0810} (2008) 091
  [arXiv:0806.1218 [hep-th]].


 \bibitem{bandos}
  I.~Bandos, H.~Samtleben and D.~Sorokin,
  ``Duality-symmetric actions for non-Abelian tensor fields,''
  arXiv:1305.1304 [hep-th].

\bibitem{wolf}
  C.~Saemann and M.~Wolf,
  ``Six-Dimensional Superconformal Field Theories from Principal 3-Bundles over Twistor Space,''
  arXiv:1305.4870 [hep-th]



\bibitem{palmer}
  S.~Palmer and C.~Saemann,
  ``Self-dual String and Higher Instanton Solutions,''
  arXiv:1312.5644 [hep-th].


\bibitem{howe}
  P.~S.~Howe, N.~D.~Lambert and P.~C.~West,
  ``The Selfdual string soliton,''
  Nucl.\ Phys.\ B {\bf 515} (1998) 203
  [hep-th/9709014].


  \bibitem{osborn}
  H.~Osborn,
  ``Solutions Of The Dirac Equation For General Instanton Solutions,''
  Nucl.\ Phys.\ B {\bf 140} (1978) 45.


  E.~Corrigan, P.~Goddard, H.~Osborn and S.~Templeton,
  ``Zeta Function Regularization And Multi - Instanton Determinants,''
  Nucl.\ Phys.\ B {\bf 159} (1979) 469.


   \bibitem{halfhet}
  G.~Papadopoulos,
  ``New half supersymmetric solutions of the heterotic string,''
  Class.\ Quant.\ Grav.\  {\bf 26} (2009) 135001
  [arXiv:0809.1156 [hep-th]].




  \end{thebibliography}
\end{document}